\documentclass[12pt]{article}
\pdfoutput=1 

\usepackage{jheppub} 
                                          
\usepackage{placeins}
\usepackage{amsmath}
\usepackage{amssymb}
\usepackage{mathtools}

\newcommand{\agt}{\gtrsim}
\newcommand{\alt}{\lesssim}
\newcommand{\bm}{\boldsymbol}

\newcommand{\beq}{\begin{equation}}
\newcommand{\eeq}{\end{equation}}
\newcommand{\bea}{\begin{align}}
\newcommand{\eea}{\end{align}}
\newcommand{\beas}{\begin{align*}}
\newcommand{\eeas}{\end{align*}}

\newcommand{\bx}{{\bm x}}
\newcommand{\by}{{\bm y}}

\newcommand{\eps}{{\varepsilon}}

\newcommand{\Tint}[1]{{\hbox{$\sum$}\!\!\!\!\!\!\!\int\,}_{\!\!\!\!\raise-0.9ex\hbox{$\scriptstyle{#1}$}}}

\definecolor{dunkelgrau}{rgb}{0.43,0.43,0.43}

\newcommand{\basispl}{
   \put(-.5,-.5){\line(1,0){1}}
   
   \put(.5,-.5){\line(0,1){1}}
   \put(.5,.5){\line(-1,0){1}}
   
   \put(-.5,.5){\line(0,-1){1}}
   }
   
   \newcommand{\basisar}{
   \put(0,-.5){\vector(1,0){0}}
   \put(.5,0){\vector(0,1){0}}
   
   \put(0,.5){\vector(-1,0){0}}
   
   \put(-.5,0){\vector(0,-1){0}}
}

   \newcommand{\revar}{
   \put(0,-.5){\vector(-1,0){0}}
   \put(.5,0){\vector(0,-1){0}}
   
   \put(0,.5){\vector(1,0){0}}
   
   \put(-.5,0){\vector(0,1){0}}
   }

\newcommand{\plaq}{\setlength{\unitlength}{.5cm}\raisebox{-.2cm}{
   \begin{picture}(1.2,1.2)(-.6,-.6)
   \basispl\basisar
   \put(-.5,-.5){\circle*{.2}}
   
   \put(-.55,-.55){\makebox(0,0)[tr]{\footnotesize $x$}}
   \put(-.55,0){\makebox(0,0)[r]{\footnotesize $\nu$}}
   
   \put(0,-.55){\makebox(0,0)[t]{\footnotesize $\mu$}}
   \end{picture}}}
   
\newcommand{\twoplaq}{\setlength{\unitlength}{1cm}\raisebox{-.5cm}{
   \begin{picture}(1.2,1.2)(-.6,-.6)
   \basispl
   \put(-.5,-.5){\circle*{.1}}
   \put(-.5,.5){\circle*{.1}}
   
   \put(.5,-.5){\circle*{.1}}
   \put(.5,.5){\circle*{.1}}
   \put(0,-.5){\circle*{.1}}
   \put(0,.5){\circle*{.1}}
   \put(.5,0){\circle*{.1}}
   \put(-.5,0){\circle*{.1}}
   \put(-.25,-.5){\vector(1,0){0}}
   \put(.25,-.5){\vector(1,0){0}}
   \put(.5,-.25){\vector(0,1){0}}
   
   \put(.5,.25){\vector(0,1){0}}
   \put(-.25,.5){\vector(-1,0){0}}
   \put(.25,.5){\vector(-1,0){0}}
   
   \put(-.5,-.25){\vector(0,-1){0}}
   \put(-.5,.25){\vector(0,-1){0}}
   
   \put(-.55,-.55){\makebox(0,0)[tr]{\footnotesize $x$}}
   \put(-.55,0){\makebox(0,0)[r]{\footnotesize $\nu$}}
   \put(0,-.55){\makebox(0,0)[t]{\footnotesize $\mu$}}
   \end{picture}}}

\newcommand{\clover}{\setlength{\unitlength}{.5cm}\raisebox{-.5cm}{
   \begin{picture}(2.4,2.4)(-1.2,-1.2)
   \multiput(-1.2,-1.2)(1.2,1.2){2}{\begin{picture}(1.2,1.2)(-.6,-.6)
   
   \basispl\basisar\end{picture}}
   \multiput(-1.2,0)(1.2,-1.2){2}{\begin{picture}(1.2,1.2)(-.6,-.6)
   \basispl\revar\end{picture}}
   
   \put(-.1,-.1){\circle*{.2}}
   \put(-.1,.1){\circle*{.2}}
   
   \put(.1,-.1){\circle*{.2}}
   \put(.1,.1){\circle*{.2}}
   
   \put(-1.2,-0.5){\makebox(0,0)[r]{\footnotesize $\nu$}}
   \put(-0.5,-1.2){\makebox(0,0)[t]{\footnotesize $\mu$}}
   \end{picture}}}

\title{\bf Non-Abelian chiral instabilities at high temperature on the lattice}

\author[a,b]{Yukinao Akamatsu,}
\author[c]{Alexander Rothkopf}
\author[d]{and Naoki Yamamoto}

\affiliation[a]{Department of Physics and Astronomy, Stony Brook University, Stony Brook, New York 11794-3800, United States}
\affiliation[b]{Department of Physics, Osaka University, Toyonaka, Osaka 560-0043, Japan}
\affiliation[c]{Institute for Theoretical Physics,  Heidelberg University, Philosophenweg 16, 69120 Heidelberg, Germany}
\affiliation[d]{Department of Physics, Keio University, Yokohama 223-8522, Japan}

\emailAdd{yukinao.akamatsu@stonybrook.edu}
\emailAdd{rothkopf@thphys.uni-heidelberg.de}
\emailAdd{nyama@rk.phys.keio.ac.jp}

\date{\today}

\abstract{We report on an exploratory lattice study on the phenomenon of chiral instabilities in non-Abelian gauge theories at high temperature. It is based on a recently constructed anomalous Langevin-type effective theory of classical soft gauge fields in the presence of a chiral number density $n_5=n_{\rm R}-n_{\rm L}$. Evaluated in thermal equilibrium using classical lattice techniques it reveals that the fluctuating soft fields indeed exhibit a rapid energy increase at early times and we observe a clear dependence of the diffusion rate of topological charge (sphaleron rate) on the the initial $n_5$, relevant in both early universe baryogenesis and relativistic heavy-ion collisions. The topological charge furthermore shows a drift among distinct vacuum sectors, roughly proportional to the initial $n_5$ and in turn the chiral imbalance is monotonously reduced as required by helicity conservation.}

\begin{document}

\maketitle

\flushbottom

\FloatBarrier

\section{Introduction}

Elucidating the properties of chiral plasmas is central to deepen our understanding of the physics of the electroweak era of the early Universe \cite{Joyce:1997uy,DiazGil:2007dy}, the quark-gluon plasma created in heavy-ion collision experiments \cite{Kharzeev:2007jp, Fukushima:2008xe}, as well as electron and neutrino media in supernovae \cite{Ohnishi:2014uea, Yamamoto:2015gzz}, to name a few. 
What distinguishes a chiral plasma from conventional plasmas is the presence of anomalous transport, such as the chiral magnetic effect (CME) \cite{Vilenkin:1980fu,Nielsen:1983rb,Alekseev:1998ds,Fukushima:2008xe}. In this particular phenomenon a current parallel to magnetic fields is induced if a chiral imbalance, characterized by the chiral density $n_5 \equiv n_{\rm R} - n_{\rm L}$, is present. When the participating gauge fields fluctuate, the CME leads to a new type of plasma instability christened the chiral plasma instability (CPI), or simply, the chiral instability \cite{Akamatsu:2013pjd, Akamatsu:2014yza}. The time evolution of phenomenologically relevant chiral plasmas exhibiting such an instability (both electroweak or quark-gluon) has not yet been fully understood, partly due to the nonlinear nature of non-abelian gauge fields. (See Refs.~\cite{Manuel:2015zpa, Buividovich:2015jfa, Hirono:2015rla} for recent attempts on \emph{Abelian} gauge theories with a chiral imbalance.) 

Take as example the the yet-to-be-settled question of whether the CME can be observed in relativistic heavy-ion collision at RHIC and LHC via event-by-event charge separation measurements \cite{Abelev:2009ac,Abelev:2012pa,Adamczyk:2014mzf}. Its solution will require a thorough quantitative understanding of anomalous transport including its physical time scales and thus the effect of the chiral instability. In this paper, we report on a first study of the nonlinear evolution and the fate of the chiral instability for high-temperature non-Abelian chiral plasmas using non-perturbative real-time lattice methods.

In a heavy-ion collision both color and electromagnetic fields are at play. The two incident nuclei traveling along the light cone can be described in the theory of the Color Glass Condensate \cite{Iancu:2003xm,Gelis:2010nm} as highly occupied gluon fields at small Bjorken $x$ and a collection of classical color sources at large $x$ \cite{McLerran:1993ni,McLerran:1993ka}. At the collision instance the interplay between the two sectors leads to the emergence of strong boost-invariant color electric and color magnetic fields, parallel to each other, spanning between the receding collision fragments \cite{Lappi:2006fp}. The fact that both fields lie parallel to each other corresponds to the presence of non-vanishing chiral charge. These initially strong and classical (or nonfluctuating) fields subsequently decay either via classical Yang-Mills evolution or via particle production facilitated by the Schwinger mechanism. There have been indications from model studies \cite{Kovner:1995ja,Lappi:2006nx} that quark production might happen quite rapidly, such that chiral fermions are present already at very early times: after a short but finite time the collision region can fragment in many subdomains of finite topological charge possessing a corresponding chiral imbalance as would be implied by the Atiyah-Singer index theorem in a Euclidean setting.%
\footnote{For a discussion of differences between real-time and Euclidean time topological charge changing processes see, e.g., Ref.~\cite{Arnold:1987zg}} The fact that we cannot cite definitive timescales in fm for these processes is a reminder of the need for more quantitative numerical studies elucidating these dynamics from first principles at early times, currently underway in several groups \cite{Berges:2013eia,Kasper:2014uaa,Gelis:2015eua}.

If the incident nuclei are viewed as classical point-sources of electromagnetic charge \cite{Tuchin:2013ie,Tuchin:2015oka}, Maxwell's equations in the form of the L\'ienard-Wiechert potentials tell us that they too generate electric and magnetic fields that lie parallel to each other. With field strengths of the order of the pion mass squared they may be capable of influencing the subnuclear composition of the collision bulk region. While the recession of the nuclei quickly diminishes such magnetic fields, the presence of quark degrees of freedom early on, i.e., their contribution to a non-equilibrium electric conductivity (for a state-of-the-art equilibrium estimate see Refs.~\cite{Ding:2010ga,Ding:2015sfa}) can potentially elongate the lifetime of these U(1) magnetic fields. The interplay between the U(1) magnetic fields and a possible chiral imbalance is a possible candidate to produce charge separation in relativistic heavy-ion collisions \cite{Kharzeev:2015znc,Huang:2015oca}. 

We are interested in the system at a point in time shortly after the initial collision, at which the external classical fields generated by the collision fragments have become negligible.  The collision region is assumed to host several distinct regions of CP-odd matter corresponding to a finite chiral density $n_5$ and an accompanying finite Chern-Simons number, or topological charge. Overall the bulk will possess vanishing chiral charge. An important aspect of this idealized setting is that helicity is at this point a conserved quantity, and so changes in the chiral imbalance have to be necessarily accompanied by opposite changes in the topological charge of the dynamical gauge fields. 

We now take one of the CP-odd domains and zoom into its interior, being interested in the physics of fluctuations of the soft gauge fields, which operate on length scales much smaller than the extend of the CP-odd domain. Hence we will take $n_5$ in the following to be spatially homogeneous from the point of view of the gauge sector. The quantum back-reaction of the chiral imbalance onto the gauge fields will then lead to some of the gauge field modes to become unstable and to exhibit rapid growth over time, the chiral instability. How these modes evolve and eventually saturate and how the presence of the chiral imbalance will change the transitions between topological sectors within the the CP-odd domain is what we wish to shed light on in the following sections.

We are able to investigate these phenomena related, to the quantum back-reaction of the chiral fermions onto the gauge field sector by deploying a recently constructed Langevin-type effective theory for soft gauge fields that captures the chiral instability \cite{Akamatsu:2014yza}. Here the label {\it soft} refers to the scale $g^2 T$, where $g$ is the coupling constant of the gauge theory and $T$ is the temperature. In the following, we will use the label {\it hard} for the scale $T$ and {\it semi-hard} for the scale $gT$.  Using classical statistical simulation techniques we numerically determine the diffusion constant for Chern-Simons number [see Eq.~(\ref{eq:rate}) for its definition], or the so called sphaleron rate. As will be shown in Fig.~\ref{Fig:SphaleronAnomalous} below, we find that the presence of $n_5$ together with the chiral instability quantitatively modifies the sphaleron rate with potential phenomenological consequences for relativistic heavy-ion collisions but also for the question of baryogenesis in the early Universe.%
\footnote{For a selection among the vast literature of studies on topology in the context of electroweak baryongenesis without explicit consideration of chiral instabilities see Refs.~\cite{McLerran:1990zh,Turok:1990in,Ambjorn:1990pu,Bodeker:1995pp,Ambjorn:1995xm,Arnold:1996dy,Ambjorn:1997jz,Moore:1997cr,   Moore:1997sn,Bodeker:1998hm,Arnold:1998cy,Bodeker:1999gx,Moore:1998zk,GarciaBellido:2002aj,Skullerud:2003ki,GarciaBellido:2003wd,Tranberg:2003gi,vanderMeulen:2005sp,D'Onofrio:2014kta}.}

The organization of the paper is as follows. In Sec.~\ref{SecCont} we review the continuum formulation of the chiral Langevin theory constructed in Ref.~\cite{Akamatsu:2014yza}. Section~\ref{SecLattice} discusses our implementation of the chiral Langevin theory in the context of classical statistical lattice simulations in thermal equilibrium and the technical challenge associated with the determination of topological charge. Our main numerical results on the chiral instability is subsequently presented in Sec.~\ref{SecNum}., while Sec.~\ref{SecConcl} is devoted to a discussion of these results and concludes the paper.

\section{Chiral Langevin theory in continuum}
\label{SecCont}

We start here with a brief summary of the Langevin-type effective theory for soft gauge fields (with momentum $k \sim g^2T$) at high temperature in the presence of a chiral imbalance \cite{Akamatsu:2014yza}. It is assumed that the system temperature is sufficiently large compared to the dynamical scale of the non-Abelian gauge theory so that $g\ll 1$. The first relevant scale below the {\it hard} scale $T$ is the {\it semi-hard} electric scale $k \sim m_{\rm D} \sim gT$ at which electric fields still propagate, but the corresponding gluons receive a thermal mass $m_D$, the Debye mass. At even lower momenta,  $k \sim g^2T$ the electric fields have become fully screened and do not propagate any longer. At this {\it soft} or magnetic scale the color magnetic fields remain as dynamical degree of freedom possessing a magnetic screening mass $m_{\rm M}\sim g^2 T$, which is generated via fully nonperturbative dynamics at this scale. 

Note first that the soft degrees of freedom of can be regarded as classical, because these modes with the momentum $k\ll T$ are highly occupied in equilibrium as
\begin{eqnarray}
n(k)=\frac{1}{e^{k/T}-1}\simeq \frac{T}{k}\gg 1\,.
\end{eqnarray}
Note also that the soft gauge fields with $A \sim gT$ and $k\sim g^2 T$ interact strongly among themselves via the classical Yang-Mills equation, because the kinetic term ($k$) and the interaction term ($gA$) in the covariant derivatives are of the same order in $g$. In addition to their self interaction, the key dynamics of the soft gauge fields arises from the color currents $\bm j_{\rm hard}$ of hard particles with the momentum $p\sim T$. The non-Abelian version of Ampere's law reads
\begin{eqnarray}
\label{eq:CYM_current}
{\bm D}\times {\bm B} = D_t {\bm E} + {\bm j}_{\rm hard},
\label{Eq:Maxw}
\end{eqnarray}
where $D_{\mu}$ is the covariant derivative. Here and below, ${\bm E}$ and ${\bm B}$ denote the color electromagnetic fields and we will often suppress color indices for simplicity.

\subsection{Langevin-type effective theory without chiral imbalance}

To make the paper self-contained, we first consider the Langevin-type effective theory for soft gauge fields without a chiral imbalance \cite{Bodeker:1998hm, Arnold:1998cy, Arnold:1999jf, Litim:1999ns,Litim:1999id}. The color current manifests itself as an effective degree of freedom, which in the leading log approximation can be expressed locally in terms of the soft gauge fields,
\begin{gather}
\label{eq:colour_current}
{\bm j}_{\rm hard}=\sigma_c {\bm E} + {\bm \zeta}, \\
\label{eq:conductivity}
\sigma_c = \frac{m_{\rm D}^2}{3\gamma} \sim \frac{T}{\ln (1/g)}\,,  \\
\label{eq:normalization}
\langle\zeta^a_i(x)\zeta^b_j(y)\rangle = 2\sigma_c T\delta^{ab}\delta_{ij}\delta^{(4)}(x-y),
\end{gather}
where $\gamma$ is the damping rate for hard gauge bosons and the coefficient $\sigma_c$ is called the color conductivity in analogy with the Ohmic law for the electric current (even though color is not a conserved number). The normalization of the right-hand side of Eq.~(\ref{eq:normalization}) is chosen such that  the noise ${\bm \zeta}$ satisfies the fluctuation dissipation theorem.

In the presence of soft gauge fields, the color current relaxes towards an Ohmic current in Eq.~\eqref{eq:colour_current} due to semi-hard scatterings with the momentum transfer $q\sim gT$ among the hard particles. The reason why semi-hard scatterings dictate the relaxation of the color current is that the effect of the semi-hard scattering on the momentum of hard particles is small as $p\sim T \gg q$, while that on their color is significant. Therefore the non-Abelian nature of the gauge fields is essential to obtain the color conductivity in Eq.~(\ref{eq:conductivity}), which allows us to approximate the color current in the local form as Eq.~\eqref{eq:colour_current} in the leading log approximation.%
\footnote{In contrast, charge conductivity of an Abelian plasma, such as the high-temperature QED plasma, is of order $T/e^2$ and the current may not be given in a local form at the scale $k\sim e^2T$.}

To proceed let us take the gauge $A^0=0$. From Eqs.~\eqref{eq:CYM_current} and \eqref{eq:colour_current} we can determine the time scale of the soft gauge fields, $\tau_{\rm soft}$. We assume that $\tau_{\rm soft} \gg 1/\sigma_c$, which will be justified in a moment. Then we can ignore the $D_t {\bm E}$ term and obtain
\begin{eqnarray}
\label{eq:Langevin1}
\sigma_c \partial_t {\bm A} = - {\bm D} \times {\bm B} + {\bm \zeta},
\end{eqnarray}
from which we have
\begin{eqnarray}
\label{eq:time_soft}
\tau_{\rm soft}\sim \frac{\sigma_c}{k^2} \sim \frac{1}{g^4 T\ln (1/g)} \gg \frac{1}{\sigma_c} \,,
\end{eqnarray}
which validates the assumption above.
By writing down a three-dimensional effective Hamiltonian of the color magnetic fields as
\begin{eqnarray}
H_{\rm eff}=\frac{1}{2}\int d^3x {\bm B}^2, 
\end{eqnarray}
one can rewrite Eq.~\eqref{eq:Langevin1} as
\begin{eqnarray}
\label{eq:Langevin2}
\sigma_c \partial_t {\bm A} = - \frac{\delta}{\delta \bm A} H_{\rm eff}[{\bm A}]+ {\bm \zeta}\,.
\end{eqnarray}
This Langevin equation \eqref{eq:Langevin2} drives the soft gauge fields towards the equilibrium distribution:
\begin{eqnarray}
P_{\rm eq}[{\bm A}] \propto \exp\left(- \frac{H_{\rm eff}[{\bm A}]}{T}\right).
\end{eqnarray}

The soft gauge fields give the dominant contribution to the Chern-Simons number diffusion rate $\Gamma$:
\begin{eqnarray}
N_{\rm CS}(t)&=&\frac{g^2}{32\pi^2}\epsilon_{ijk}
\int d^3x \left(
F^a_{ij}A^a_k - \frac{g}{3}f_{abc}A^a_iA^b_jA^c_k
\right)\,,\\
\label{eq:rate}
\Gamma &=&\lim_{V\to \infty} \lim_{t\to \infty}
\frac{\langle\left(N_{\rm CS}(t) -N_{\rm CS}(0)\right)^2\rangle}{Vt}\,,
\end{eqnarray}
where $f_{abc}$ is the structure constant. The $N_{\rm CS}$-changing processes require the condition, $g^2 AB R^3\sim g^2 B^2 R^4 \agt 1$, where $R\sim 1/k$ is the wave length of the mode.
The energy for such a mode is $E_A \sim B^2 R^3 \agt 1/(g^2R)$, which should satisfy $E_A \alt T$ so as not to be suppressed in the thermal environment. This gives the condition $1/k\sim R\agt 1/g^2T$, but gauge fields with $k\ll g^2T$ are suppressed due to the magnetic screening mass $m_{\rm M}\sim g^2T$. Therefore, the soft gauge fields $k\sim g^2T$ contribute dominantly to the Chern-Simons number diffusion rate. Since we know that the soft gauge fields evolve at time scale $\tau_{\rm soft} \sim 1/[g^4 T\ln(1/g)]$ [see Eq.~(\ref{eq:time_soft})], the Chern-Simons number diffusion rate is parametrically estimated to be 
\begin{eqnarray}
\Gamma \sim \frac{k^3}{\tau_{\rm soft}} \sim g^{10} T^4 \ln (1/g).
\end{eqnarray}
The scales of the soft gauge fields and their power counting are summarized in Tab.~\ref{tab:softscales1}.

\begin{table}
\centering
\label{tab:softscales1}
\begin{tabular}{c|c}
\ \ \ \ \ \ \ \ \ \ \ \ \ \ \ & \ \ Counting in $g$ \ \ \\ \hline
$k$ & $g^2T$  \\
$\omega$ & $g^4 T \ln(1/g)$ \\
$\bm A(x)$ & $gT$ \\ 
$\bm E(x)$ & $g^4 T^2 \ln(1/g)$ \\
$\bm B(x)$ & $g^3 T^2$ \\
$\Gamma$ & $g^{10} T^4 \ln(1/g)$ \\ \hline
\end{tabular}
\caption{Scales of the soft gauge fields.} 
\end{table}

\subsection{Langevin-type effective theory with chiral imbalance}
So far, we have not considered the nature of the hard modes in the plasma, which couple to the soft gauge fields through the current $\bm j_{\rm hard}=\sigma_c \bm E + \bm \zeta$. In fact the hard modes with the momentum $p\sim T$ consist of gauge fields and fermions that interact with the soft gauge fields. Without the imbalance of chiral fermions, they would affect the Debye screening mass $m_{\rm D}$ only via the number of colors and flavors. However, in the presence of chiral fermions, the $N_{\rm CS}$-changing processes in the soft sector induce changes of their chirality through the anomaly relation,
\begin{eqnarray}
\partial_{\mu} j^{\mu 5} = \frac{N_f g^2}{4\pi^2}\bm E\cdot \bm B.
\end{eqnarray}
Therefore, the Langevin equation needs to be amended to take into account nonzero chiral charge \cite{Akamatsu:2014yza}. The evolution of the chiral imbalance $N_5=N_{\rm R} - N_{\rm L}$  itself is driven by the change in the topological charge 
\begin{eqnarray}
\label{eq:n5_anomaly}
\dot N_5(t) &=& \frac{N_f g^2}{4\pi^2}\int d^3x \bm E\cdot \bm B = -\dot N_{\rm CS}(t),\\
\label{eq:n5_mu5} n_5(t) &\equiv& \frac{N_5(t)}{V} 
= N_c N_f \left[\frac{1}{3\pi^2}\left(\mu_5^3 + 3\mu_5 \mu^2\right) + \frac{1}{3}T^2\mu_5 \right] \,,
\end{eqnarray}
where we denote the chiral chemical potential by $\mu_5=\mu_5(t)$. Equation (\ref{eq:n5_anomaly}) means that total helicity $N_5 + N_{\rm CS}$ is conserved as a function of time. The back reaction of the chiral fermions on the gauge sector is implemented via the non-Abelian version of a chiral magnetic current proportional to the color magnetic field ${\bm B}$,
\begin{eqnarray}
\label{eq:modified_current}
\bm j_{\rm hard}=\sigma_c \bm E + \sigma_{\rm anom} \bm B + \bm \zeta, \ \ \
\sigma_{\rm anom} = \frac{N_f g^2 \mu_5}{4\pi^2}.
\end{eqnarray}
In the form of Eq.~\eqref{eq:Langevin2}, the latter modification is compactly expressed by
\begin{eqnarray}
\label{eq:Langevin3}
H_{\rm eff}[\bm A] &=&  \frac{1}{2}\int d^3x {\bm B}^2 + 2N_f \mu_5 N_{\rm CS},\\
\sigma_c \partial_t {\bm A} &=& - \frac{\delta}{\delta \bm A} H_{\rm eff}[{\bm A}]+ {\bm \zeta}. \nonumber
\end{eqnarray}
This derivation of the modified Langevin theory utilizes a power counting of $g$ assuming $\mu_5\sim T$ in the parametric sense.
Note also that all the terms of $\bm j_{\rm hard}$ are of the same order $g^5 T^3$.

This modified effective theory \cite{Akamatsu:2014yza} possesses the following two new properties which the original one \cite{Bodeker:1998hm, Arnold:1998cy, Arnold:1999jf, Litim:1999ns,Litim:1999id} does not have:
\begin{enumerate}
\item Chern-Simons number diffusion receives back reaction from the fermion sector.
\item Soft gauge fields possess an unstable mode in the presence of finite $\mu_5$ as argued in Ref.~\cite{Akamatsu:2013pjd}.
\end{enumerate}
Let us consider the time scales of the associated phenomena. The order of the right-hand side of Eq.~\eqref{eq:n5_anomaly} is $g^{10}T^4\ln (1/g)$, and so the time scale for $\mu_5 \sim T$ is thus estimated to be 
\begin{eqnarray}
\tau_{\mu_5}\sim \frac{1}{g^{10} T\ln(1/g)}\,.
\end{eqnarray}  
The time scale of the instability is obtained from Eq.~\eqref{eq:Langevin3} as%
\footnote{
Strictly speaking, the soft gauge fields are strongly interacting ($k \sim gA \sim g^2T$) so that the linear analysis is not applicable close to equilibrium. It is applicable only when the system starts from $A\ll gT$.}
\begin{eqnarray}
\tau_{\rm inst} \sim \frac{\sigma_c}{k^2} \sim \frac{1}{g^4 T \ln(1/g)}\,.
\end{eqnarray}
The unstable mode carries the Chern-Simons number so that it reduces the chiral imbalance $\mu_5$. Due to the separation of the time scales, $\tau_{\rm inst}\ll \tau_{\mu_5}$, the instability develops and saturates due to nonlinear interactions among the soft gauge fields during the time in which $\mu_5$ only changes very little. 

At the longer time scales $\tau_{\mu_5}$, the fermion chiral charge decreases towards its equilibrium value. The questions here are
\begin{itemize}
\item What is the fate of the instability? How do the nonlinear interactions saturate the instability?
\item What is the fate of the fermion chiral imbalance? In particular, what is the fraction of $N_5$ and $N_{\rm CS}$ eventually?
\end{itemize}
In order to understand these questions which are highly nonperturbative and nonlinear, we now turn to numerical lattice simulations in the following.

\section{Lattice discretized effective theory and numerical setup}
\label{SecLattice}

Let us discuss in detail the numerical implementation of the real-time evolution of the soft gauge fields in the background of chiral fermions. The qualitatively new aspect of this work is the inclusion of anomalous affects related to the quantum back-reaction of a dynamically evolving chiral imbalance onto the gauge sector, described effectively by the term proportional to the ${\bm B}$ field in Eqs.~\eqref{Eq:Maxw}, \eqref{eq:n5_anomaly}, and \eqref{eq:modified_current}. Note that we are solving the full Hamiltonian equations of motions and do not neglect the $D_t \bf E$ term. While implementing the Hamilton dynamics of the gauge fields is straightforward, the treatment of topological charge on the lattice constitutes a challenge. Since the lattice definition of the Chern Simons number suffers from the fact that it cannot be related to a total derivative, it is known that spurious fluctuations on the level of the lattice spacing contaminate its numerical evaluation (see, e.g., Ref.~\cite{Moore:1996qs}). In order to robustly identify changes in the topology of our system we combine several techniques, developed in both classical statistical Yang-Mills simulations, as well as Euclidean lattice QCD over the last two decades. 

\subsection{Classical Yang-Mills on the lattice}
\label{SecYMLat}

Let us start with classical Yang-Mills theory on the lattice in the absence of fermionic degrees of freedom. A common approach to the derivation of its real-time Hamilton dynamics is to start from Wilson's formulation \cite{Wilson:1974sk,Klassen:1998ua} in terms of an action on a finite hypercubic space-time grid with lattice spacings $a_s$ and $a_t$ and volume $N_s\times N_t$. The gauge degrees of freedom are link variables $U_\mu(\bx,t)$ that are related to the gauge fields $A^a_\mu(\bx,t)$ in the adjoint representation via
\begin{align}
U_\mu(\bx,t)={\rm exp}[ia_\mu g A_\mu^a (\bx,t) T^a], \\
T^a=\frac{\lambda^a}{2},\quad{\rm Tr}[T^aT^b]=\frac{1}{2}\delta^{ab},\quad {\rm Tr}[T^a]=0\,,
\end{align}
where $\lambda^a$ represents either the Pauli matrices in the case of SU(2) $(a=0,1,2)$ or the Gell-Mann matrices in the case of SU(3) $(a=0,\ldots,7)$. The appropriately rescaled quantities $T^a$ denote the generators of the gauge group. 

We use a shorthand notation for the lattice spacing $a_{\mu=0}=a_t$ and $a_{\mu=i}=a_s$, where summation over indices is not implied. From the gauge links one can construct the smallest gauge invariant object, the plaquette $U_{\mu\nu}$, which is related to the field strength tensor $F_{\mu\nu}$ as
\begin{align}
U_{\mu\nu}&=U_\mu(x)U_\nu(x+\hat{\mu})U_\mu^\dagger(x+\hat{\nu})U^\dagger_\nu(x)={\rm exp}[ig a_\mu a_\nu F^a_{\mu\nu} T^a].
\end{align}

As is common practice, we consider only the simplest formulation of the Wilson action 
\begin{align}
\nonumber S[U]=&-\frac{2a_sN_c}{a_tg^2}\sum_{\bm x} \sum_{i} \Big[ \frac{1}{2N_c}\Big({\rm Tr}[U_{0i}]+{\rm Tr}[U^\dagger_{0i}]\Big)-1\Big] \\
&+\frac{2a_tN_c}{a_sg^2}\sum_{\bm x} \sum_{i<j} \Big[ \frac{1}{2N_c}\Big({\rm Tr}[U_{ij}]+{\rm Tr}[U^\dagger_{ij}]\Big)-1\Big]\,,
\end{align}
to derive the equations of motion. 
In order to arrive at an equation of motion of Hamiltonian form, we take the established route of fixing to temporal gauge $A_0(\bx,t)=0$, i.e., $U_0(\bx,t)=1$ after evaluating the stationarity condition
\begin{align}
 \frac{\delta S[U]}{\delta A^a(x)}=0 \,,
\end{align}
with
\begin{align}
 \frac{\delta U_\mu(\bx,t)}{\delta A^b_\nu(y)}= iga_\mu T^a\delta_{\nu\mu}\delta_{\bx \by}U_{\mu}(\bx,t),\quad  \frac{\delta U^\dagger_\mu(\bx,t)}{\delta A^b_\nu(y)}= -iga_\mu U^\dagger_{\mu}(\bx,t) T^a\delta_{\nu\mu}\delta_{\bx \by} \,.
\end{align}
From the functional derivative with respect to the spatial fields and using a historic choice of normalization $E^a_k=F^a_{0k}ga_s/(2\sqrt{N_c})$ we arrive at the Hamilton-like classical equation of motion,
\begin{align}
 \partial_tE^b_k(\bx,t)&=-\frac{1}{\sqrt{N_c a_s^2}} {\rm Im Tr}\Big[T^bU_k(\bx,t)\sum_{j\neq k}(S^\sqsubset_j+S^\sqsupset_j)\Big]\,,  \label{Eq:EFldEOM} \\
 \dot{U}_k(\bx,t)&=\Big(iga_sF_{0k}^a(\bx,t)T^a\Big)U_k=\Big(i2\sqrt{N_c}E^a_k(\bx,t)T^a\Big)U_k\,,
\end{align}
where the second line follows from a Taylor expansion of the temporal plaquette. The staples $S^\sqsubset_j$ and $S^\sqsupset_j$ are defined such, that when multiplied with the link $U_j$ from the right or left respectively, lead to the closed backward and forward plaquette. 

The particular form of these equations invites the use of a second order accurate ${\cal O}(dt^2)$ leap-frog scheme for their implementation. The electric fields are first evolved one half time step via a simple Euler update and then $E$ and $U$ are subsequently updated based on their shifted counterpart. While the time derivative for $E$ is naively discretized the update of the links is implemented exactly via matrix exponentiation. For SU(2) this is achieved via the handy relation,
\begin{align}
 \theta=\sum_a E^aE^a, \, \kappa=a_t \sqrt{ N_c  \theta }, \quad {\rm exp}\big[2i\sqrt{N_C}E^a_kT^a\big]= {\bf 1} {\rm cos}(\kappa)+ \frac{2{\rm sin}(\kappa)}{\sqrt{\theta}}E^a_kT^a\,,
\end{align}
while for SU(3) we would resort to the fast and robust matrix diagonalization of Ref.~\cite{Kopp:2006wp} based on Cardano's analytic formula.

Classical Yang-Mills fields represent a constrained system, a fact reflected in the functional derivative in the $A^0$ direction leading to a relation between fields that does not involve a partial derivative in time, the so called Gauss constraint,
\begin{align}
\nonumber &G(\bx,t)\equiv \sum_{i}\big( E^a_i(\bx,t) T^a - U_i^\dagger(\bx-\hat{i},t) E^a_i(\bx-\hat{i},t)T^a U_i(\bx-\hat{i},t)\big).
\end{align}
If we start from an initial field configuration that respects the Gauss constraint, the application of Eq.~\eqref{Eq:EFldEOM} will preserve this property. Only the accumulation of rounding errors gradually weakens the constraint, which can be conveniently monitored via the global quantity $\sum_{b,\bx}{\rm Tr}\Big[T^b G(\bx,t)\Big]$. 

Note that while we have started from spatial and temporal links on a hypercubic spacetime grid, we have arrived at a formulation of electric fields and spatial links on individual three-dimensional time-slices, which evolve via (continuous) time derivatives in Eq.~\eqref{Eq:EFldEOM}. The Hamiltonian corresponding to these classical equations of motion reads
\begin{align}
 \bar{H}[E,U]=\frac{1}{a_t}\Big[ S_{\rm el}[E]-S_{\rm mag}[U]\Big]= \frac{2N_c}{g^2 a_s} \sum_{\bm x} \Big[ \sum_{i} a_s^2E^a_iE^a_i-\frac{1}{N_c}\sum_{i<j}\Big({\rm ReTr}[U_{ij}]-N_c\Big)\Big].
\end{align}

\begin{figure}
\centering
 \includegraphics[scale=0.55]{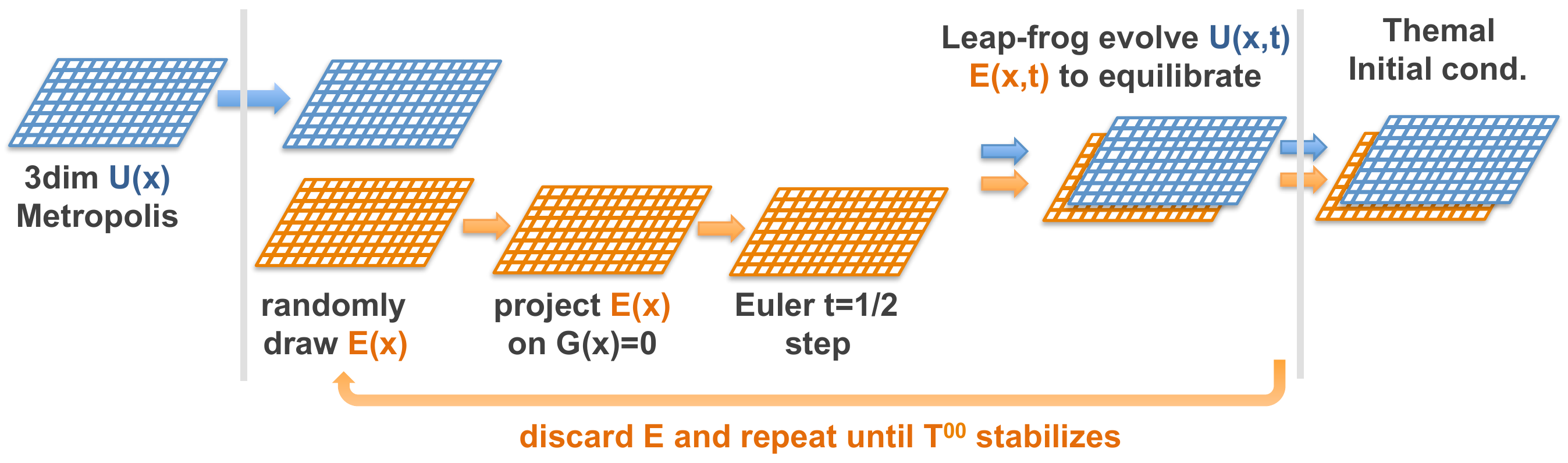}
 \caption{Schematic view of the generation of a thermal initial condition as discussed in the main text.}
 \label{Fig:InitCond}
\end{figure}

In this study we wish to investigate the physics in a thermal equilibrium setting for which we note that the Boltzmann weight can be written in natural units as
\begin{align}
 P[E,U] \propto {\rm exp}\Big( -\frac{1}{T} \bar{H}[E,U] \Big) \equiv {\rm exp}\Big( - \underbracket{\frac{2N_c}{g^2a_sT}}_{\beta_L} H[E,U] \Big).
\end{align}
The properties of a thermal configuration are thus solely determined by the lattice parameter $\beta_L=\frac{2N_c}{g^2a_sT}$. In practice simulations are carried out at a fixed $\beta_L$ using $a_s=g=1$.

To generate thermal initial configurations we follow the well established strategy of Refs.~\cite{Grigoriev:1988bd,Ambjorn:1990pu,Ambjorn:1997jz} with the implementation of Refs.~\cite{Laine:2013lia,Laine:2013apa}, which is described in detail in the following and schematically depicted in Fig.~\ref{Fig:InitCond}. It relies on the fact that the electric fields enter quadratically in the Hamiltonian.
\begin{enumerate}
 \item Preconditioning of spatial links: Using the Wilson plaquette part of the Hamiltonian a standard three-dimensional multihit Metropolis Monte-Carlo implementation $(N_{\rm hits}=10\times400)$ brings the spatial links from either hot- or cold-start towards an ensemble member of the Markov chain representing  $P[U]\propto {\rm exp}\big[-S_{\rm mag}[U]\big]$. This is not yet the correct thermal distribution but the links are already closer to the correct ones than, e.g., a randomly chosen configuration.
 \item The electric field distribution in thermal equilibrium does not explicitly depend on the gauge fields and $P[E]\propto{\rm exp} \big[ -\frac{2N_Ca}{g^2T}\sum_{\bx,j,a} E_j^a(\bx)E_j^a(\bx)\big]$ and hence we may start out by drawing each field entry separately from a Gaussian distribution 
 \begin{align}
  E^a_j=\xi^a_j,\quad \langle \xi^a_j \rangle =0,\quad \langle \xi^a_j\xi^b_k \rangle= \sigma^2 \delta_{ab}\delta_{jk},\quad \sigma=\frac{1}{2\beta_La_s^2}\, .
 \end{align}
 \item The electric fields drawn such contain both physical and unphysical modes, as only the former obey the Gauss law. Hence we need to project out those fields that are perpendicular to the constrained directions. A particularly clever way to do so is via the iterative prescription of Ref.~\cite{Moore:1996bn}, where one computes for each $E_k(\bx)$ the Gauss constraint at $\bx$ and $\bx+\hat{k}$ and carries out the following update
 \begin{align}
 E^a_k(\bx) \to E^{a\prime}_k(\bx)=  2 {\rm Tr}\Big( T^a \big[ \tilde{\kappa} \big( U_k(\bx) G(\bx) U^\dagger_k(\bx) G(\bx+\hat{k}) - G(\bx) \big)  - E^a_kT^a \big]\Big) \,.
 \end{align}
 The choice of $\tilde{\kappa}=0.12$ has been found to lead to fast and stable convergence, allowing us to enforce the Gauss constraint down to $\sum_{\bx} G(\bx)=10^{-15}$.
 \item After bringing the projected electric fields forward by $a_t/2$ via a naive Euler step, they are evolved together with the gauge links according to the Hamilton dynamics of Eq.~\eqref{Eq:EFldEOM} to mutually equilibrate for $(N_{\rm eq.\,steps}=300)$.  After the evolution phase, the current electric field is discarded and one repeats from step 2 by drawing another round of Gaussian $E_k^a$. This cycle is itself repeated $N_{\rm cycl}=40$ times and at the same time the approach of the energy $T^{00}=F^{0\mu}F^{0\mu}+  F_{\mu\nu}F^{\mu\nu}$ towards a time independent value is monitored.
\end{enumerate}
For the steps involving pseudo-random numbers we deploy L\"uscher's ranlux algorithm \cite{Luscher:1993dy}, as it is proven to yield statistically independent chains if started from differing seeds.

Naive classical thermal equilibrium is ill-defined in the continuum due to the Rayleigh-Jeans divergence,%
\footnote{The incorporation of hard-thermal loops as means to make the continuum limit well-defined is discussed, e.g., in Ref.~\cite{Bodeker:1999gx}.} 
nevertheless on the lattice it is regulated by the cutoff $\pi/a_s$. Since in such a scenario all modes up to the cutoff are occupied we have to expect that lattice artifacts do significantly influence the simulated physics. The physics of the semi-hard scale at $m_{\rm D}\sim gT$ indeed suffers from lattice artifacts, as can be seen from the explicit $a_s$ dependence of the lattice regularized Debye mass parameter \cite{Bodeker:1995pp,Arnold:1997yb,Watson:1939,Glasser:2000} computed in lattice perturbation theory
\begin{align}
 m_{\rm D}^2=2g^2TN_c\frac{\Sigma}{4\pi a_s},\quad \Sigma=\Gamma^2\left[\frac{1}{24}\right]\Gamma^2\left[\frac{11}{24}\right]\frac{\sqrt{3}-1}{48\pi^2}\,.
\end{align}
It is the physics at scales $\sim g^2T$ and below that is considered to be reproduced quantitatively within the classical lattice theory. The physics of topological transitions is among the phenomena that can be studied in this approach.

\subsection{Topological charge on the lattice}

We have seen in Sec.~\ref{SecCont} that the gauge field topology is intimately related to the dynamics of the chiral fermions via the anomaly and vice versa. Hence we need robust means to identify how the topology of the gauge field changes over time. In this study we use the naive continuum integral formula,
\begin{align}
 \frac{dN_{\rm CS}}{dt}= \frac{g^2}{64\pi^2} \int d^3x \, F^a_{\mu\nu}(\bx)\tilde{F}^{\mu\nu}_a(\bx)\,,
 \label{Eq:ContNCS}
\end{align}
where the ${F}^{\mu\nu}$ is defined from a clover-type approximation \cite{GarciaPerez:1993ki} summing over the field strength components extracted from four neighboring plaquettes 
\begin{align}
 F^{\mu\nu}= -\frac{1}{4}\frac{i}{ga_\mu a_\nu}\sum_\square {\rm log}\left(   \clover \right)\, .
\end{align}
Instead of the matrix logarithm one may also use the simple relation $ F^{\mu\nu} \approx \frac{1}{2}{\rm Re}[U^{\mu\nu}]-\frac{1}{N_C}{\rm Tr}[U^{\mu\nu}]$. Since we need also access to the temporal plaquettes at this point we will keep a copy of the fields at the previous time step in memory even though they are not needed to evolve Eq.~\eqref{Eq:EFldEOM}. This however means that only the two backward plaquettes are used to define $F^{0k}=-F^{k0}$.  While in this study we deploy the standard clover approximation to the field strength tensor, in principle one can improve on the definition \cite{Moore:1996wn}, which leads to an expression for ${\bm E}\cdot {\bm B}$ to which lattice spacing artifacts contribute only at ${\cal O}(a_s^4)$.

As has been discussed in the literature in great detail (see, e.g., Ref.~\cite{Moore:1998swa}), the naive evaluation of Eq.~\eqref{Eq:ContNCS} on a lattice suffers from the fact that it is not related to a total derivative. It hence becomes susceptible to fluctuations at small length scales that are unrelated to the topology. Indeed it was shown that modifying a single link \cite{Moore:1996qs} can change the lattice discretized value of $\frac{dN_{\rm CS}}{dt}$. In other words, difference in topological charge between two configurations, 
\begin{align}
 N_{\rm CS}(t)-N_{\rm CS}(0)= \frac{g^2}{64\pi^2} \sum_t a_t  \sum_x a_s^3 \, F^a_{\mu\nu}\tilde{F}^{\mu\nu}_a \, ,
 \label{Eq:TopChrgLat}
\end{align}
depends in general on the path in field space (here along real-time) one takes to connect the two. In a finite temperature setting we should note that $N_{\rm CS}$ does not necessarily has to take on integer values as it does in vacuum, as thermal fluctuations allow it to explore the topological sector it currently resides in. In the presence of these ambiguities, the only truly well-defined piece of information is what topological vacuum sector the current field configuration belongs to. It is this integer value we set out to distill in the following.

Note that focusing on the topological sector alone introduces other conceptual difficulties. It has been argued \cite{Arnold:1996dy} that gauge fields can spend significant time in the vicinity of a sphaleron configuration at half-integer topological charge, moving slightly from one sector to the neighboring one and back over timescales of $1/(g^2T)$. If we naively counted the number of transitions between sectors this would then overestimate how topology actually diffuses over time. In our case we use the formula 
\begin{align}
 \Gamma = \lim_{t\to\infty} \frac{\gamma(t)}{t} = \lim_{t\to\infty} \Big(\frac{2N_C}{4\pi\beta_L}\Big)^{-5} \frac{\big[ \langle (N_{\rm CS}(0)-N_{\rm CS}(t))^2\rangle -\langle N_{\rm CS}(0)-N_{\rm CS}(t)\rangle^2   \big] }{N^3 t}\,,
 \label{Eq:ScaledSphaleron}
\end{align}
for the sphaleron rate $\Gamma$, where the rapid changes around a sphaleron configuration will average out in the late time limit.

In order to numerically extract the change between topological sectors another possible path might lie in devising a Minkowski space implementation of the overlap operator and to measure the number of exact chiral zero modes, which at least in the Euclidean domain are related to the change in topology by the Atiyah-Singer index theorem. Another strategy laid out by Moore and Turok is the slave-field method \cite{Moore:1997cr}, which explicitly constructs a representation of the topology of the gauge group. 

A more practical way, which has been deployed widely in classical-statistical lattice simulations and which under the moniker of Wilson flow \cite{Luscher:2010iy} has recently captured also the interest of the Euclidean-time lattice QCD community is dissipative \textit{cooling}. It acts on the spatial links of the theory and loosely speaking damps away their high frequency modes. Namely, a copy of the current-time links $U(\bx)$ is evolved in a fictitious cooling time $\tau$ according to
\begin{align}
\partial_\tau A^a_j=-\frac{\delta H_{\rm cool}}{\delta A^a_j},\quad  \frac{\partial U}{\partial \tau} = \frac{\partial A^a_j}{\partial \tau} \frac{\delta}{\delta A^a_j} U_k = -\frac{\delta H_{\rm cool}}{\delta{A^a_j}}iga_sT^aU_k\,. \label{Eq:Cooling}
\end{align}
In the case of four-dimensional Euclidean lattice QCD $\tau$ represents the flow-time. Since this procedure is formulated in a gauge invariant fashion, we know that it cannot simply modify field modes outside of a certain momentum shell but also has to influence the low lying modes in a certain fashion. Indeed, in the context of Euclidean Yang-Mills theory it has been shown that cooling with the naive Wilson plaquette Hamiltonian reduces the size of any instanton currently present on the configuration. Eventually it makes the topological object small enough that it cannot be resolved on the lattice and its contribution to $N_{\rm CS}$ will vanish. Namely, with the naive plaquette Hamiltonian one can easily overcool a configuration and hence miss transitions. In response a number of studies were devoted to constructing classical actions that lead also to growth of instantons, so as to balance the overall behavior. In this study we choose to use the improved Wilson action 
\begin{equation}
H_{\rm imp.\,cool}(\eps)=\frac{4-\epsilon}{3}\sum_{x,\mu,\nu}{\rm Tr}\left(1-\plaq\right)
+\frac{\epsilon-1}{48}
\sum_{x,\mu,\nu}{\rm Tr} \left(1-\twoplaq\right)\quad \,.
\label{eq:epsac1}
\end{equation}
with $\epsilon=0$ introduced in Ref.~\cite{GarciaPerez:1993ki} and discussed in, e.g., Refs.~\cite{deForcrand:1997sq,GarciaPerez:1998ru}. In practice it gives a much smoother behavior during cooling, i.e., the step size in cooling time can be chosen much coarser than in the case of the naive single plaquette. While it has been suggested \cite{GarciaPerez:1998ru} to monitor the deviation of each individual link during cooling to avoid overcooling we do not implement this strategy here. 

To be able to obtain consistently cooled configurations, we furthermore deploy the fourth order Runge-Kutta update proposed in Ref.~\cite{Luscher:2010iy} to implement the cooling of the form $\partial_\tau U(\tau)=Z(U(\tau))U(\tau)$, given by
\begin{align}
 \nonumber Y_0=U(\tau),\quad  Y_1={\rm exp}\left(\frac{1}{4}Z_0\right)Y_0\,, \quad Y_2={\rm exp}\left(\frac{8}{9}Z_1-\frac{17}{36}Z_0\right)Y_1\, ,\\ 
  U(\tau+a_t)={\rm exp}\left(\frac{3}{4}Z_2 -\frac{8}{9}Z_1+\frac{17}{36}Z_0\right)Y_2\,
\end{align}
with $Z_i=\epsilon Z(Y_i),\,i=0,1,2$.

As was proposed first in Ref.~\cite{Moore:1998swa} one might consider an additional coarsening step at intermediate cooling time, in which one replaces two neighboring links with the product of these two links, effectively reducing the number of points on the lattice by factor $2^{D=3}$. The idea behind this is that after having carried out sufficient cooling in the first place the field modes that resolve individual lattice spacings are already highly suppressed and thus the blocked lattice will contain essentially the same information as before. While in the literature it has been suggested to use more elaborate schemes, where also staples around the coarsened links are considered in this study we refrain from such an additional smearing step. The practical benefit of coarsening is that any subsequent cooling proceeds much faster. We have checked in several cases that we end up in the same vacuum sector with and without cooling.

In summary, at each step in real-time we prepare a copy of the gauge links, cool them for a certain flow-time, carry out one blocking before finally cooling towards the point at which the behavior of $N_{\rm CS}(t)-N_{\rm CS}(0)$ stabilizes to unit steps. Since even with the improved action overcooling can still occur we have to empirically determine the appropriate cooling depth for the lattices at hand. Note that this dissipative approach should not be confused with the recently introduced gauge cooling in the context of complex Langevin simulations \cite{Seiler:2012wz,Aarts:2013uxa}, as it constitutes more than a simple gauge transformation and actually changes the value of the classical action.

Note also that since we here directly evaluate $N_{\rm CS}$ from $F^{\mu\nu}$, which is defined from plaquettes of spatial links on neighboring time-slices, the explicit electric fields of the hamilton dynamics do not enter. In turn, besides cooling the links there is no need to change the electric fields themselves. For comparison purposes we have implement the electric field cooling from Ref.~\cite{Ambjorn:1997jz} which did however not reproduce our results, a comparison with the original code was unfortunately not possible anymore.

\subsection{Effective gauge dynamics in the presence of chiral fermions}

Now that we have spelled out how the gauge fields evolve classically and how to determine their topology we can implement Eqs.~\eqref{eq:n5_anomaly} and \eqref{eq:modified_current} on the lattice. While the update of the gauge links remains unchanged the equation of motion for the electric field changes due to the presence of the Ohmic current, the corresponding thermal noise, and the dissipationless anomalous current,
\begin{align}
\label{Eq:AnomalousEOM}
 \partial E^b_k(\bx,t) =&-\frac{1}{\sqrt{N_c a_s^2}} {\rm Im Tr}\Big[T^bU_k(\bx,t)\sum_{j\neq k}(S^\sqsubset_j+S^\sqsupset_j)\Big] \\
 \nonumber&-\sigma_cE^b_k(\bx,t) -\frac{ga_s}{2\sqrt{N_C}}\sigma_{\rm anom}B^a_k(\bx,t) + \frac{ga_s}{2\sqrt{N_C}}\xi^b_k(\bx,t)\,,  
\end{align}
with
\begin{align}
 \langle \xi^a_j(\bx,t)\rangle=0, \quad \langle  \xi^a_j(\bx,t) \xi^b_k(\by,t')\rangle=\frac{4\sigma_cN_C}{g^2 a_t a_s^4 \beta_L}\delta_{\bx\by}\delta_{tt'}\delta_{ab}\delta_{jk} 
\end{align}
and $\sigma_{\rm anom}=\frac{N_fg^2\mu_5(t)}{4\pi^2}$. We explicitly implement the stochastic evolution and do not resort to heat-bath updates \cite{Moore:1998swa} nor the electric field refresh method \cite{Moore:2010jd} previously deployed in the literature. 

At this point we need an explicit expression for the magnetic field, for which we take $B_i=\frac{1}{2}\epsilon_{ijk}F^{jk}$ (i.e., $B_0=-F_{12}$, $B_1=-F_{20}$, $B_2=F_{01}$)
with $F^{jk}$ obtained from the same clover approximation used in the determination of the topological charge.

The evolution of the chiral imbalance $n_5(t)$ is driven by changes in the topology and we naively discretize the continuum equations \eqref{eq:n5_anomaly}. Since no spatial dependence of $n_5(t)$ is considered here its derivative can be related to the change in the topological charge 
\begin{align}
 \partial_t n_5(t) = \frac{1}{a_s^3 N^3}\frac{N_f g^2}{4\pi^2}\int d^3x {\bm E}(\bx)\cdot{\bm B}(\bx) = \frac{2 N_f}{a_s^3 N^3} \frac{dN_{\rm CS}}{dt}\,,
\end{align}
consistent with the requirement of helicity conservation. The chiral chemical potential $\mu_5(t)$ which is needed to evolve $E^b_k$ in Eq.~\eqref{Eq:AnomalousEOM} is taken as the unique solution of the third oder polynomial relation Eq.~\eqref{eq:n5_mu5}.

\begin{figure}
\centering
 \includegraphics[scale=0.6]{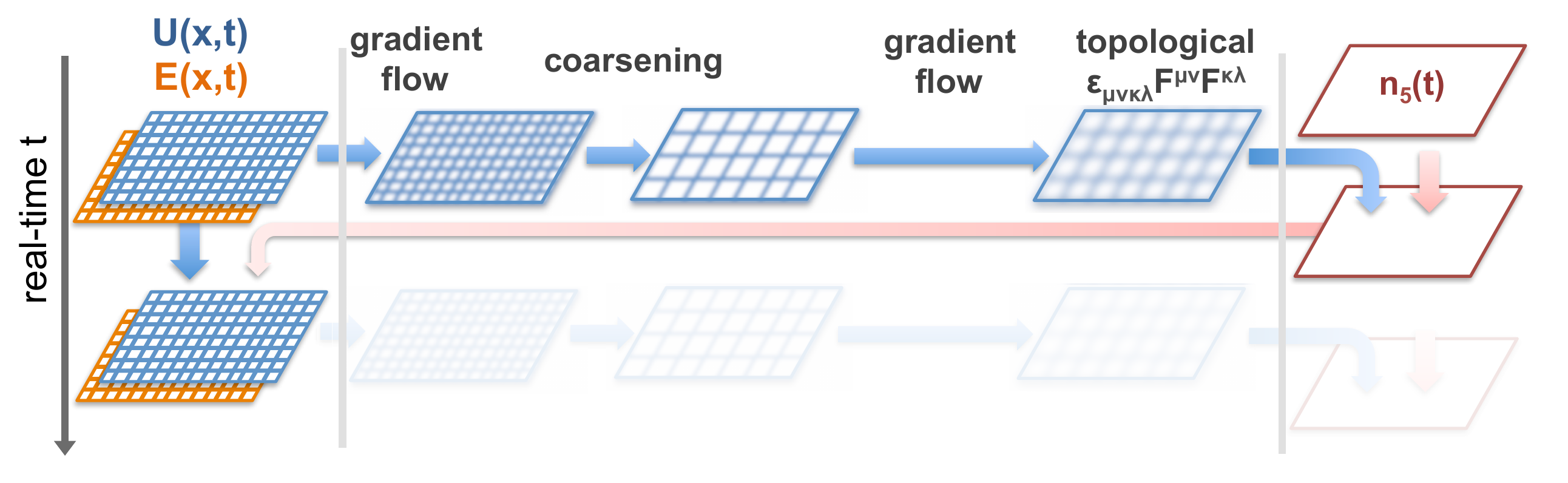}
 \caption{Schematic view of the evolution of the soft gauge fields $E^a_k(\bx,t)$ and $U_k(\bx,t)$ in the presence of a dynamical chiral imbalance $n_5(t)$ coupled effectively via an anomalous current. At each step in the leap-frog-like real-time evolution we prepare a copy of the gauge links and cool them, including a single coarsening step, in order to compute the topological contribution to $\frac{dN_{\rm CS}}{dt}\propto F^{\mu\nu}\tilde{F}_{\mu\nu}$. The change in $N_{\rm CS}$ enters the update of $n_5(t)$ which in turn via $\mu_5(t)$ influences the update of the electric fields at this step. The cooled copy of the configuration is kept in memory for one more time-step, since the electric part of $F^{\mu\nu}$ requires the evaluation of temporal plaquettes.}
 \label{Fig:AnomEvol}
\end{figure}

\section{Numerical simulations and their discussion}
\label{SecNum}

With the lattice discretized formalism in place we are ready to embark on the investigation of the effective gauge dynamics in non-Abelian SU(2) Yang-Mills theory\footnote{As we expect no qualitative differences between the gauge groups SU(2) and SU(3) our choice of the former is solely due to reduce the computational burden. Our code has been designed to be capable of handling also the gauge group SU(3).} in the presence of a chiral imbalance with $N_f=2$ flavors. Let us start first with a cross-check, whether our numerical implementation is able to reproduce the well-established results for the Yang-Mills sector. The first test concern the evolution of the topological charge in a thermal ensemble of electric fields and gauge links propagating via pure Hamilton dynamics, as was originally investigated in Ref.~\cite{Moore:1998swa}. The second testing ground is the driven dynamics of color fields in the presence of an Ohmic current and stochastic noise, for which in Ref.~\cite{Moore:2010jd} the sphaleron rate has been determined. 

We choose our lattice geometry according to the insight gained in Ref.~\cite{Moore:2010jd}. It showed that the sphaleron rate in both SU(2) and SU(3) can be reliably determined on three-dimensional lattices with an extend of fifteen points, hence we choose $N=20$. In order to not having to deal with too rough field configurations we furthermore choose the lattice coupling $\beta_L=8$ with $a_s=1,g=1$, which corresponds to the choice of $aN_Cg^2T=1$ in Tab.~1 of Ref.~\cite{Moore:2010jd}. In their study this choice of parameters already produced a result for the sphaleron rate within $1 \sigma$ of the continuum extrapolated result. We set the time discretization of the Hamilton dynamics fine enough so that the same $a_t$ can be used in all runs, even in the presence of fast dynamics, expected in the presence of a chiral imbalance. A posteriori we found that $a_t=0.0375a_s$ provides an appropriate amount of temporal resolution and at the same time allows conservation of magnetic and electric energy on the single percent level.

\subsection{Thermal Yang-Mills dynamics}

\label{SecYMTh}

\begin{figure}
 \hspace{-0.0cm}\begin{minipage}{6in}
  \centering
  \raisebox{-0.5\height}{\includegraphics[scale=0.48]{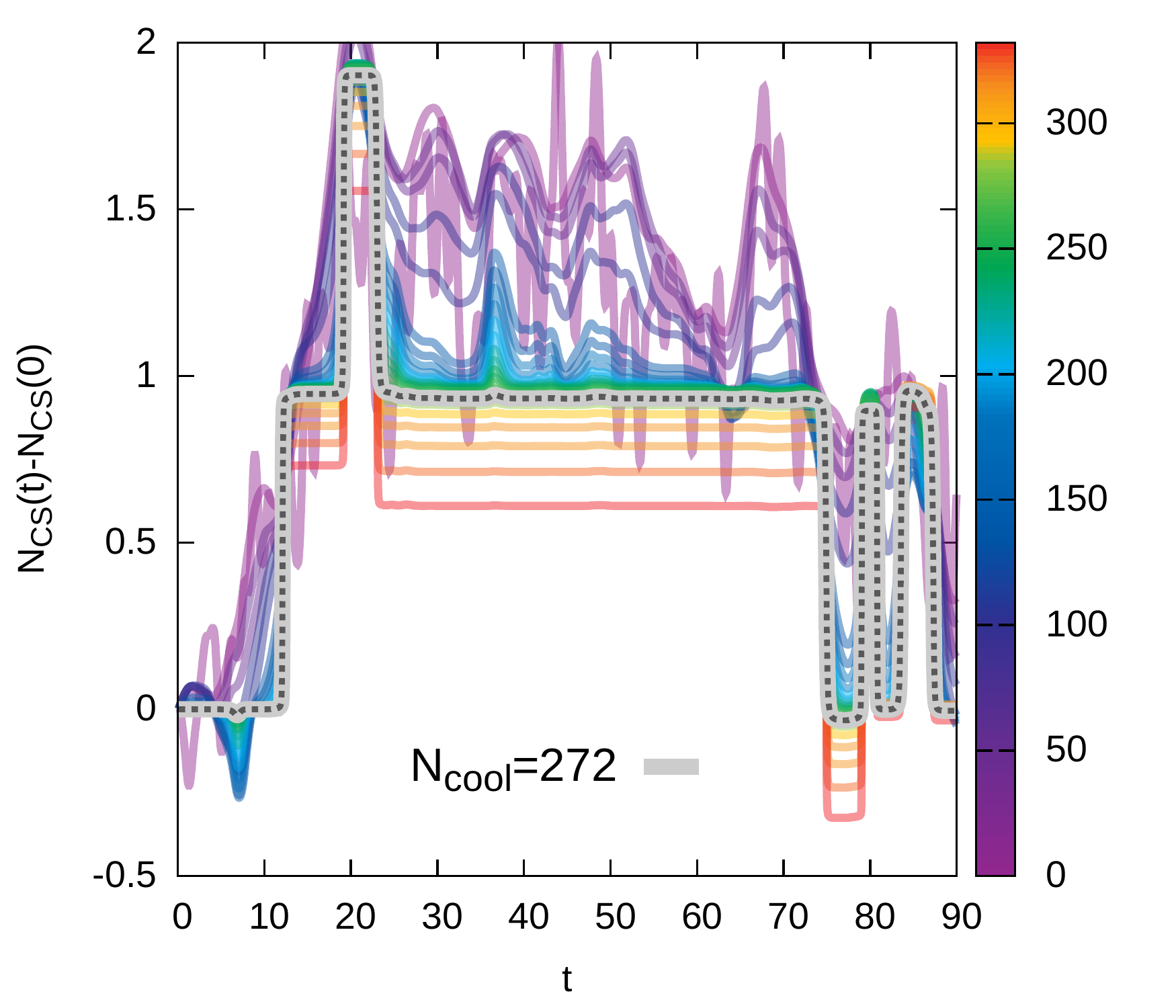}}
  \hspace*{-0.2in}
  \raisebox{-0.5\height}{\includegraphics[scale=0.315]{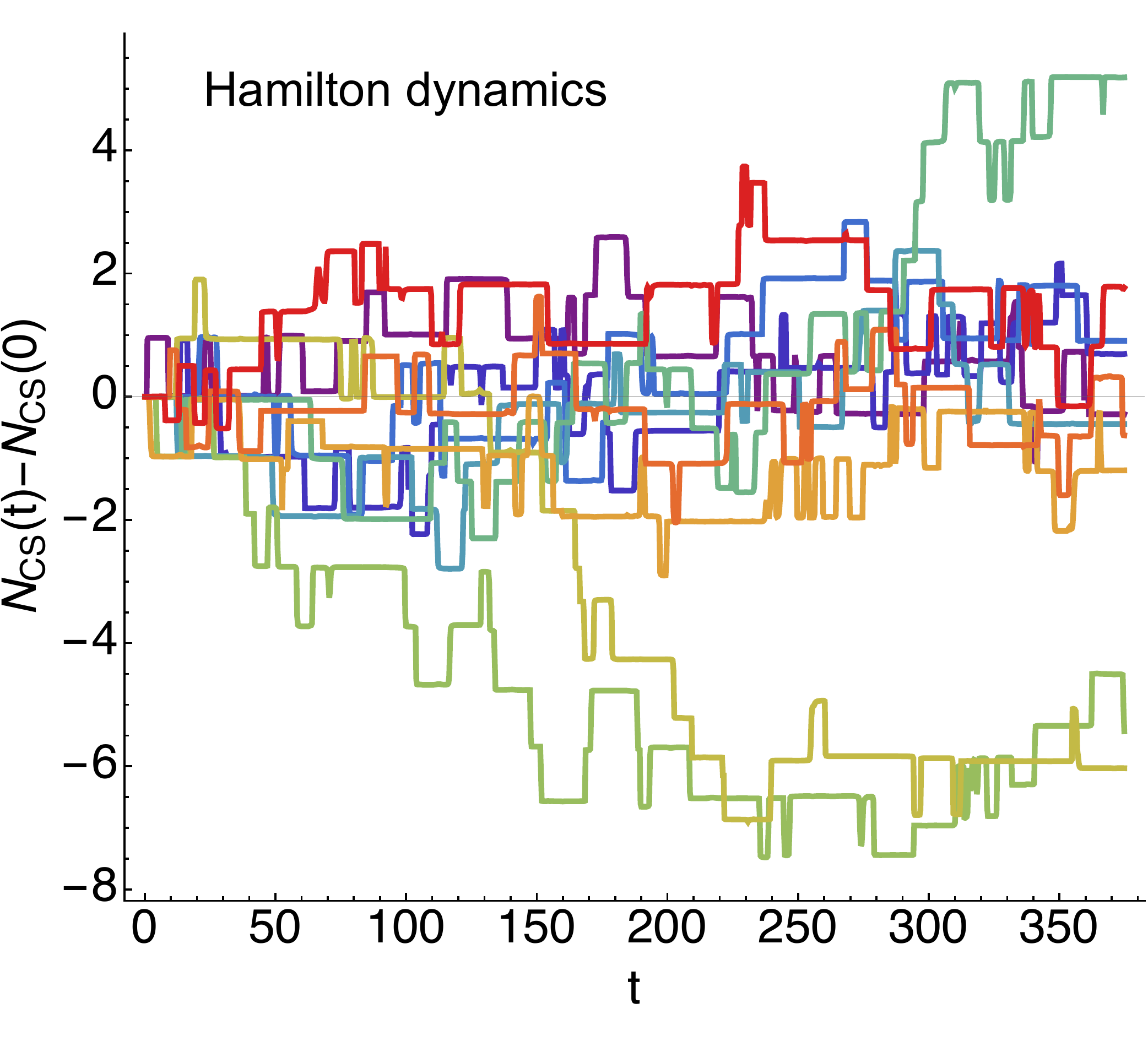}}
\end{minipage} 
 \caption{Numerical results of pure Yang-Mills theory: (Left) Estimates for $N_{\rm CS}(t)-N_{\rm CS}(0)$ from evaluating Eq.~\eqref{Eq:TopChrgLat} on a single configuration using different values for the cooling depth as given by the color coding of the individual curves. A single coarsening step is included for those curves that correspond to cooling beyond $100$ steps of $d\tau=0.15$. The optimal cooling curve $L_{\rm cool}=272\,d\tau$ is overlayed in gray. Note that further (over-)cooling significantly reduces the the height of already the first step away from unity; (Right) Early time behavior of $N_{\rm CS}(t)-N_{\rm CS}(0)$ after cooling, starting from a selection of ten different thermal initial configurations. Well separated transitions between individual vacuum sectors are observed. At several instances, rapid oscillations in $N_{\rm CS}(t)-N_{\rm CS}(0)$ around a sphaleron configuration occur (e.g., orange curve $t=250-275$), which according to Ref.~\cite{Arnold:1996dy} are related to damping effects of order ${\cal O}(1/g^2)$.  These do not adversely affect our estimate of the sphaleron rate, as Eq.~\eqref{Eq:ScaledSphaleron} does not count individual crossings but averages them out.}
 \label{Fig:NCSEvolHam}
\end{figure}

Starting from thermal initial field configurations obtained by the prescription laid out in Sec.~\ref{SecYMLat}, we evolve $E^a_k(\bx)$ and $U_k(\bx)$ via Eq.~\eqref{Eq:EFldEOM}. Our goal is a robust determination of $N_{\rm CS}(t)-N_{\rm CS}(0)$ for which it is necessary to set up an appropriate cooling procedure. The use of the fourth-order Runge-Kutta scheme allows us to use a relatively large cooling step size of $d\tau=0.15a_s$. It furthermore turns out that a single coarsening after $N_{\rm coarse}=100$ cooling steps does not adversely affect the determination of the current topological sector. In order to avoid overcooling we perform the computation of the topological charge for different cooling depths starting with the same initial condition, as shown on the left of Fig.~\ref{Fig:NCSEvolHam}. The appropriate cooling depth $L_{\rm cool}=N_{\rm cool}d\tau$ is chosen such that $N_{\rm CS}(t)-N_{\rm CS}(0)$  has approached the unit step function. As one can clearly see further cooling eventually reduces the unit step behavior and will ultimately remove all structure. With the optimal setting of $N_{\rm cool}=272$ for the pure Hamilton dynamics, we proceed to measure the topological charge on $N_{\rm conf}^{\rm Ham}=40$ ensemble realizations, a selection of which is shown on the right of Fig.~\ref{Fig:NCSEvolHam}.\footnote{The fact that we are fully cooling to the vacuum at each $a_\tau=0.0375a_s$ time step makes this simulation resource intensive, even on the relatively small $20^3$ lattices used here.}

\begin{figure}
\centering
 \includegraphics[scale=0.36]{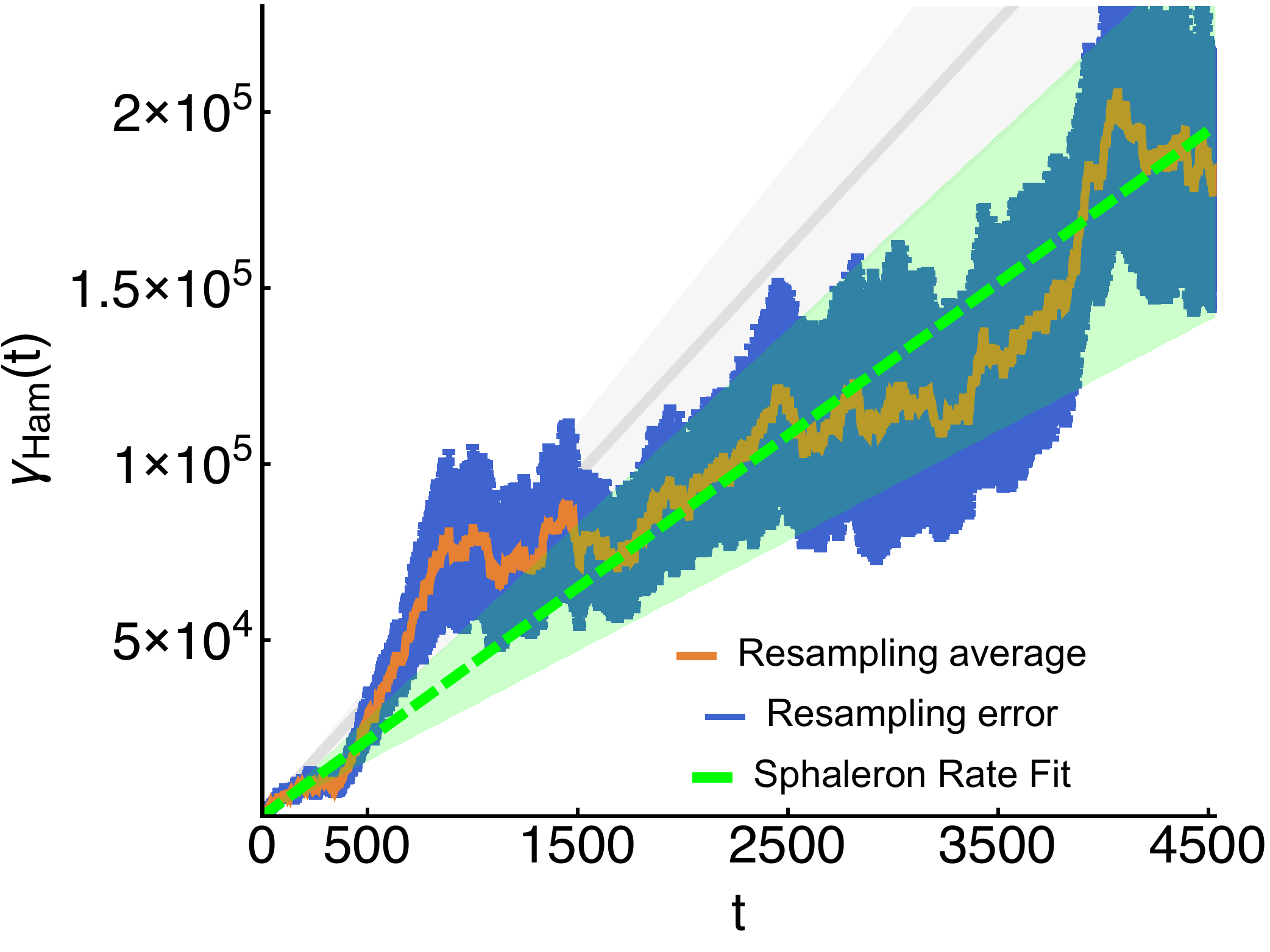}
 \includegraphics[scale=0.36]{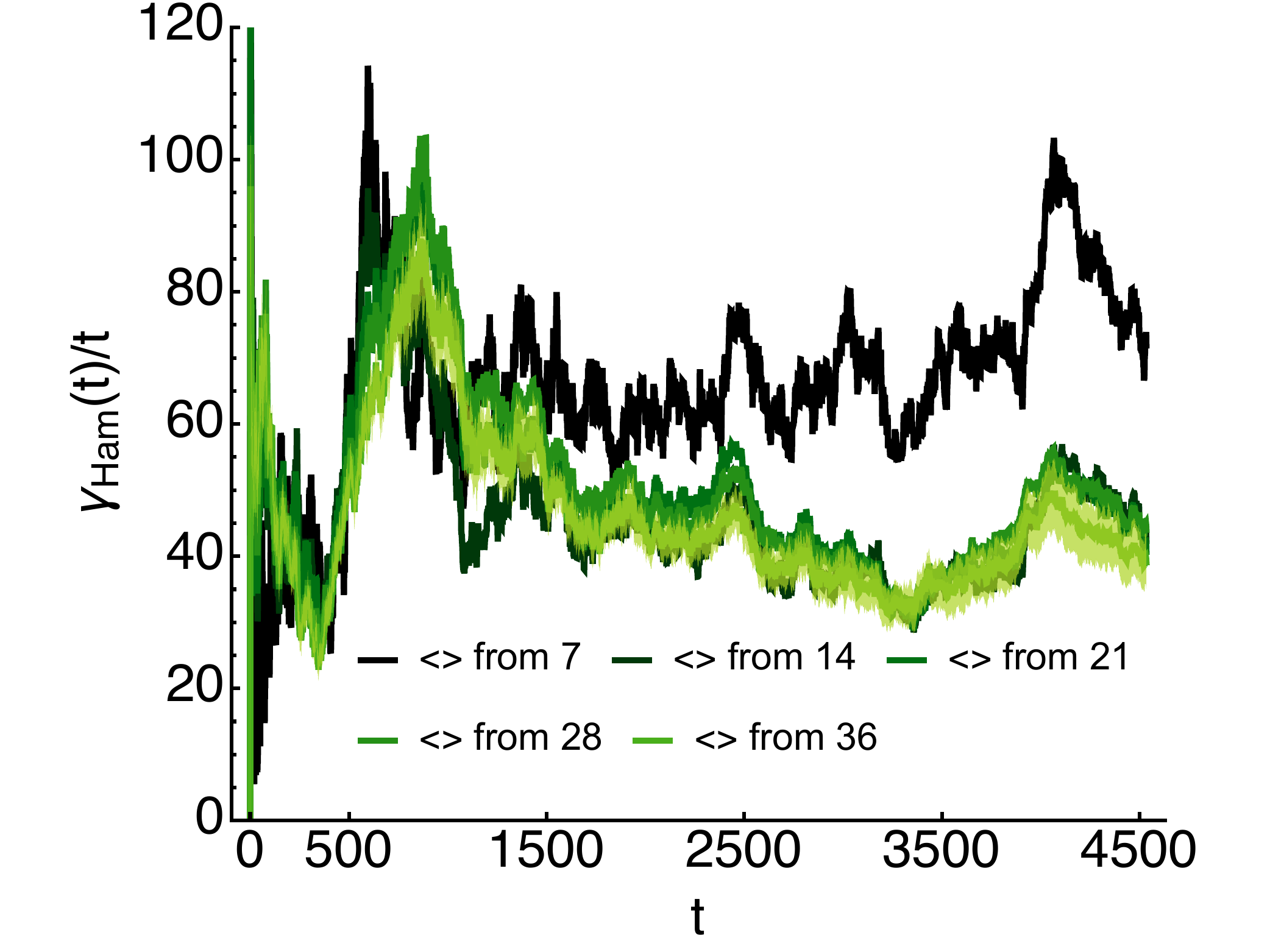}
 \caption{Numerical results of pure Yang-Mills theory: (Left) Estimating the sphaleron rate $\Gamma$ from the slope of $\gamma(t)$. We resample twenty times by drawing a subset of twenty ensemble realizations from which the average (orange) and variance (blue) of $\gamma(t)$ follows. Assuming a diffusive process we fit with a linear function without intercept (green). The gray curve denotes the value for $\Gamma=64.5$ observed in Ref.~\cite{Moore:1998swa}. (Right) The behavior of $\gamma(t)/t$ which will asymptote to the constant $\Gamma$ at late times. Different curves correspond to a different number of ensemble realizations used. Note that more than ten runs are required obtain convergence.}
 \label{Fig:SphaleronHamConv}
\end{figure}

The sphaleron rate $\Gamma$ is obtained from the appropriately rescaled variance of the topological charge $\gamma(t)$ defined in Eq.~\eqref{Eq:ScaledSphaleron}. The variance of a stochastic variable is itself a fluctuating quantity with an uncertainty we need to estimate. To this end we carry out resampling of  $\gamma(t)$ by repeatedly drawing a random selection of twenty of the forty ensemble realizations we computed numerically. This procedure leads to the blue error bands in the left panel of Fig.~\ref{Fig:SphaleronHamConv}.  

The sphaleron rate can be estimated in two similar ways. Both assume that the topological charge diffuses, i.e., its variance will grow linearly with time. In the left panel of Fig.~\ref{Fig:SphaleronHamConv} we fit the values of $\gamma(t)$ using a straight line without intercept. Fixing the function at the origin is a quite restrictive choice, which leads to relatively small error estimates when changing the upper end of the fitting interval. To estimate the inherent systematic error we assign error bars on the slope, such that the meandering orange average lies within $1\sigma$ at late times $t>2000$,

\begin{align}
 \Gamma^{\rm Ham}=44.5\pm12, 
 \label{Eq:SphRateYM}
\end{align}
to be compared to $\Gamma^{\rm Ham}_{[74]}=64.5\pm 9.7$ in Ref.~\cite{Moore:1998swa}.

\noindent The error bands of our estimate overlap with those of the published result from Ref.~\cite{Moore:1998swa}. We note however that our average value lies systematically lower than $\Gamma^{\rm Ham}_{[74]}$. Three possible reasons for this difference exist, the first lies in the cooling prescription, the second in the operator used to define ${\bm E}\cdot{\bm B}$ and the third in a different number of configurations used in the ensemble averaging. Instead of the calibrated cooling of Ref.~\cite{Moore:1998swa} with standard Wilson action, we here use the gradient flow from the improved action to directly reach the vacuum in each time step. When measuring the field strength tensor on the other hand we deploy the standard clover approximation, while Ref.~\cite{Moore:1998swa} is based on an improved operator. Last but not least the number of ensemble realizations used in the estimate influences the value of $\Gamma$. While Ref.~\cite{Moore:1998swa} does not explicitly mention the number of ensembles used, in an older study \cite{Ambjorn:1997jz} to which it refers, 20 to 60 realizations have been generated. 

In the right panel of Fig.~\ref{Fig:SphaleronHamConv} we show the values of $\gamma(t)/t$ computed from a different number of ensemble members. It seems necessary to include at least twenty of them before convergence to the correct late time result is possible. Fitting a constant to this quantity at $t>2000$ constitutes the second way of estimating $\Gamma$. We obtain a similar value as in \eqref{Eq:SphRateYM} consistent within error bars. If $\Gamma$ is obtained from a smaller number of ensemble members it shows slightly larger values and thus lies closer to the results reported in Ref.~\cite{Moore:1998swa}. 

\subsection{Langevin-type effective theory without chiral imbalance}
\label{SecYMEFT}

\begin{figure}
\centering
 \includegraphics[scale=0.33]{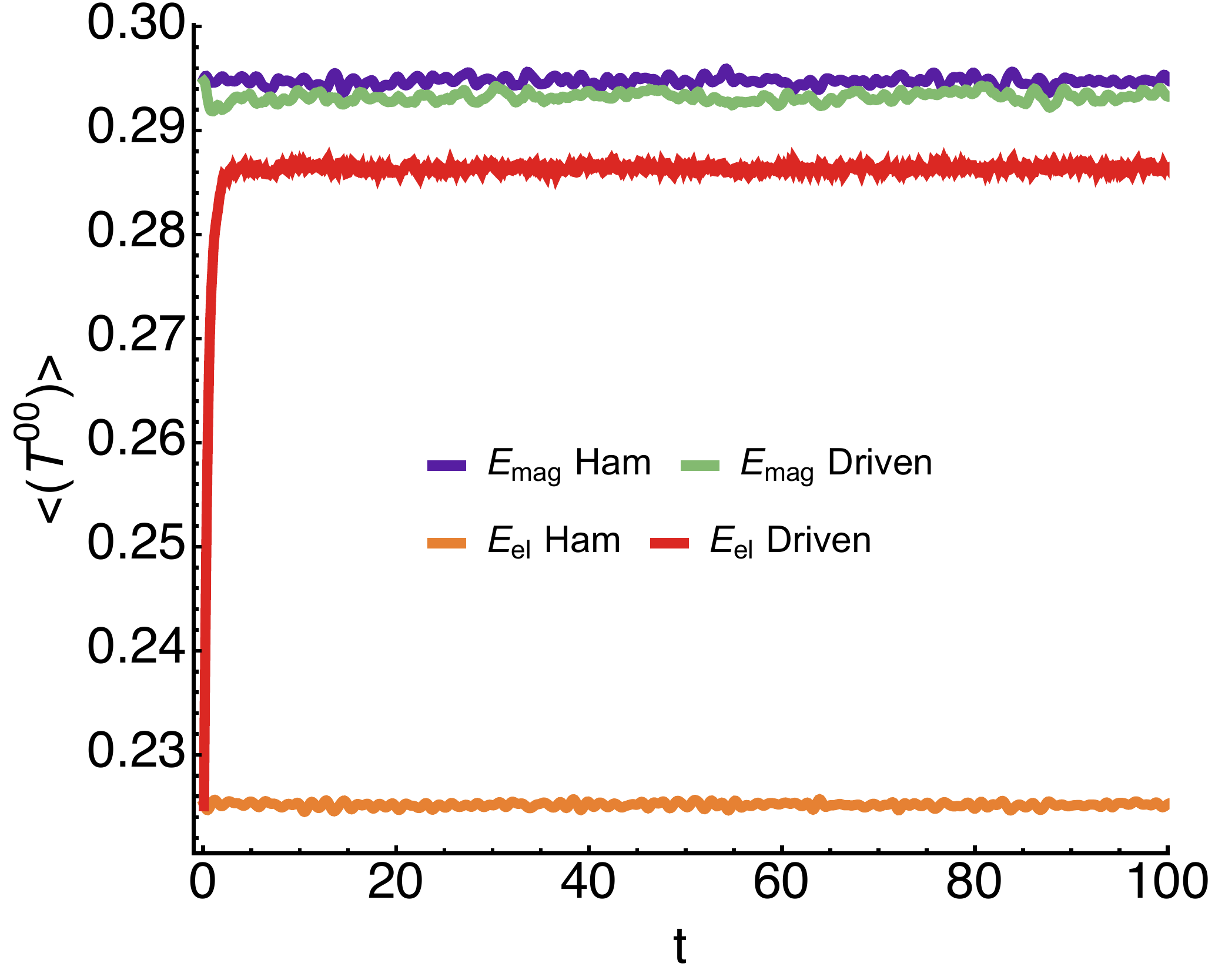}
 \caption{Numerical results of nonchiral Langevin theory: Magnetic and electric energy density $T^{00}$ of the pure Yang-Mills system compared to Yang-Mills with dissipatively coupled fermions. By construction the energy density in the magnetic sector does not deviate significantly between the two cases, while the presence of the stochastic noise leads to a marked increase in $T^{00}_{\rm el}$ before settling to a new steady state at around $t\simeq 10$. Note that $T^{00}_{\rm el} < T^{00}_{\rm mag}$ is always fulfilled.}
 \label{Fig:EnergyDrivenHam}
\end{figure}

The next step is to investigate a scenario where hard modes are coupled to the gauge fields via an Ohmic current and stochastic noise as discussed recently, e.g., in Ref.~\cite{Moore:2010jd}. Starting from the same initial thermal configurations as in Sec.~\ref{SecYMTh} we evolve the electric fields via Eq.~\eqref{Eq:AnomalousEOM}, neglecting the anomalous term at this stage. The value of the color conductivity is set to $\sigma_c=1$, as its influence can be simply absorbed into a rescaling of time \cite{Moore:2010jd}. The interplay between dissipation and the driving noise by construction leaves the energy content of the magnetic sector unchanged, which is fulfilled in our implementation within statistical errors, as can be seen form the two top curves in Fig.~\ref{Fig:EnergyDrivenHam}. There is a slight systematic shift of the energy to lower values in the driven case, which we take as contribution to our systematic uncertainties.

The electric fields on the other hand show a pronounced difference in energy reflected in the two lower curves in Fig.~\ref{Fig:EnergyDrivenHam}. The energy in the electric sector exhibit a sharp rise at early times before settling at a stationary value at $t\simeq10$.  Note that in accordance with the assumptions in the derivation of the effective dynamics (see Tab.~\ref{tab:softscales1}) the electric energy never grows beyond its magnetic counterpart $E_{\rm el}<E_{\rm mag}$. The initial rise in $E_{\rm el}$ is not particular to our numerical implementation, as the stochastic update deployed in Ref.~\cite{Moore:2010jd} also leads to a similar increase. It tells us that if we are interested in investigating possible fast dynamics from anomalous effects in the following, we should first let the driven system equilibrate to its stationary state before switching on a chiral imbalance.

\begin{figure}
\centering
 \includegraphics[scale=0.4]{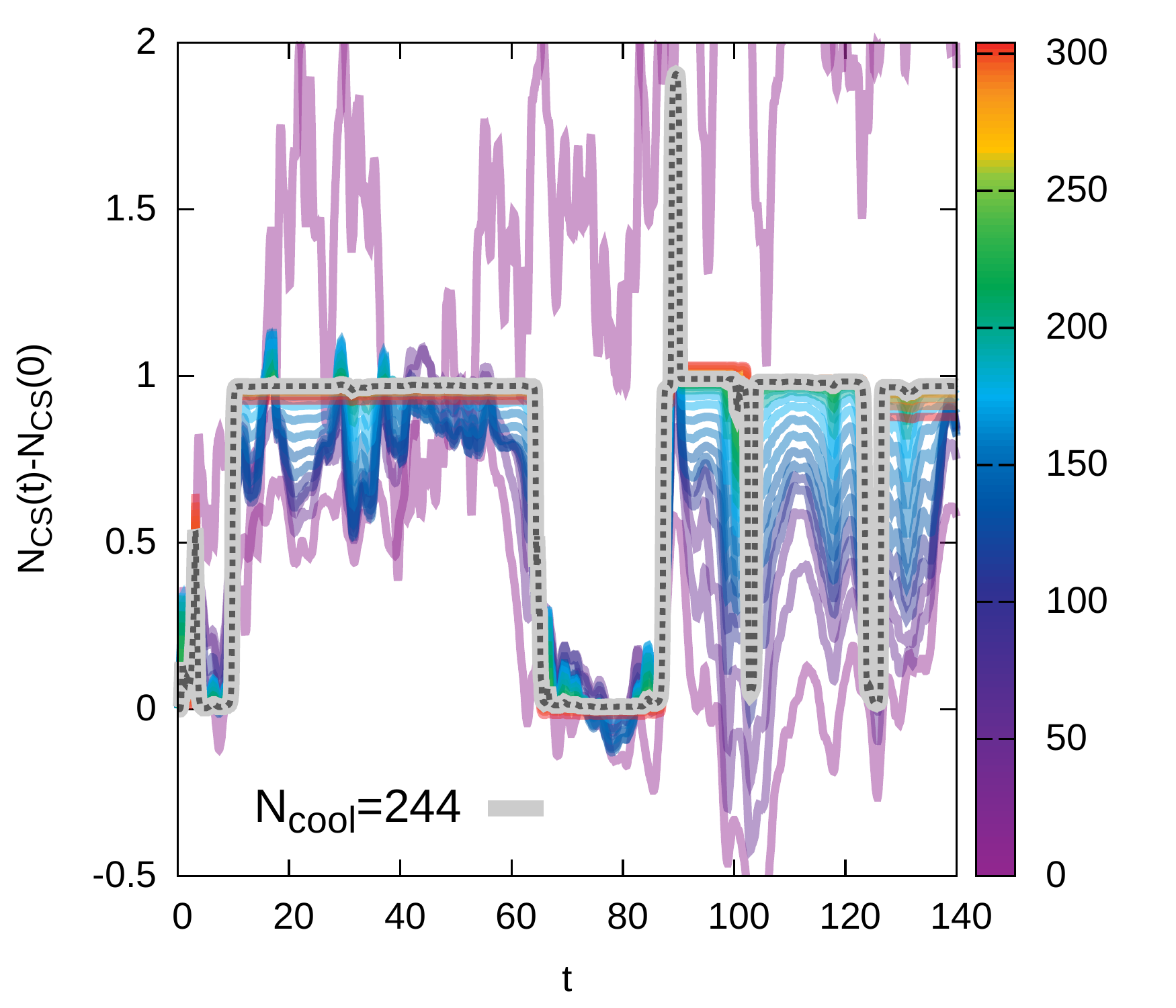}
  \includegraphics[scale=0.315]{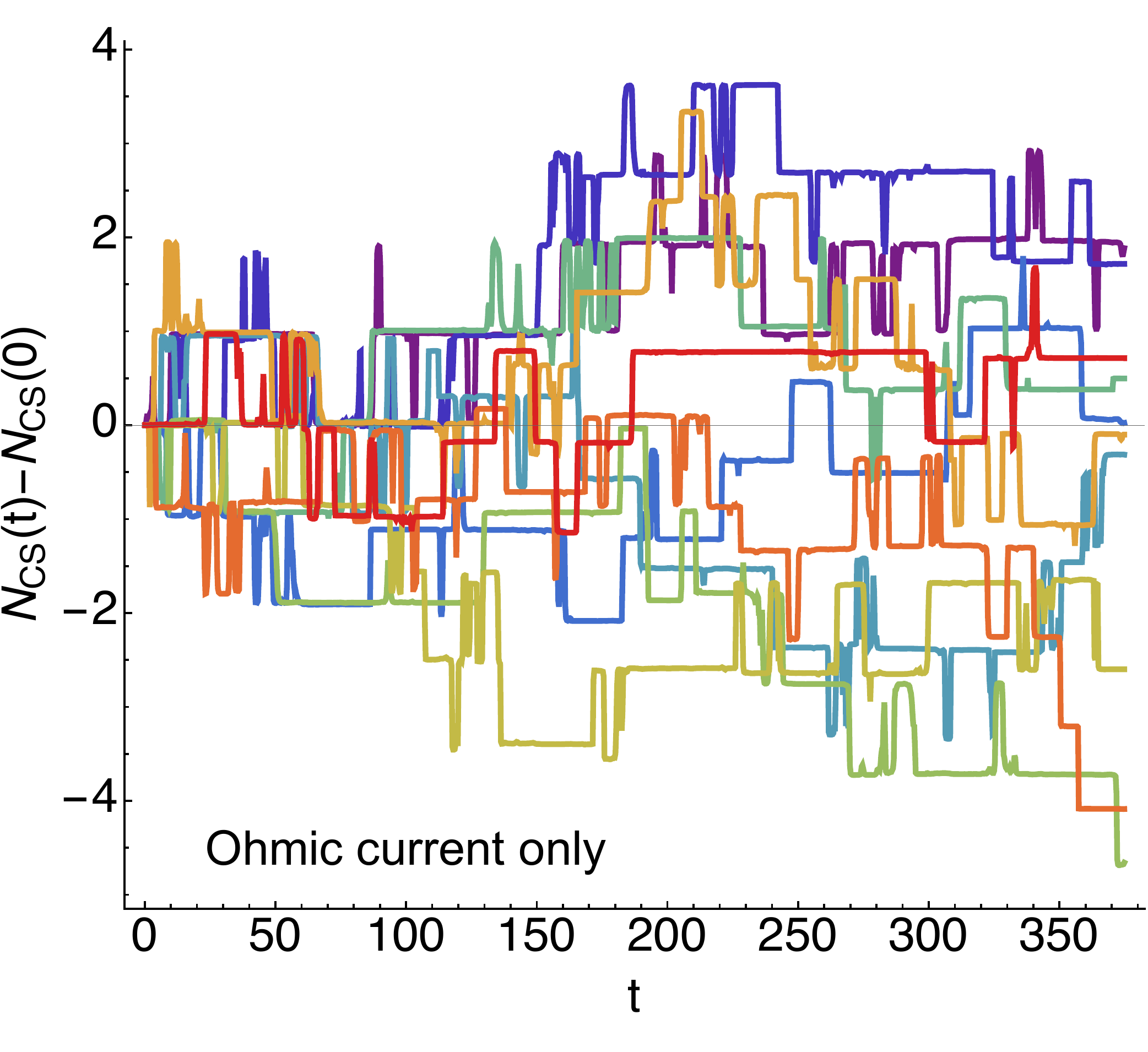}
 \caption{Numerical results of nonchiral Langevin theory: (Left) $N_{\rm CS}(t)-N_{\rm CS}(0)$ evaluated on a single configuration using different values for the cooling depth as given by the color coding of the individual curves. A single coarsening step is included for those curves that correspond to cooling beyond $100$ steps of $d\tau=0.15$. The optimal cooling curve $L_{\rm cool}=244\,d\tau$ is overlayed in gray (dashed); (Right) Early time behavior of $N_{\rm CS}(t)-N_{\rm CS}(0)$ after cooling, starting from a selection of ten different thermal initial configurations. Well separated transitions between individual vacuum sectors are observed. }
 \label{Fig:NCSDriven}
\end{figure}

The very different fluctuation content compared to pure Yang-Mills theory also requires us to choose a different cooling scheme for the robust determination of $N_{\rm CS}(t)-N_{\rm CS}(0)$. With the same $d\tau=0.15$ as before and a single coarsening after $N_{\rm coarse}=100$ steps, we find from Fig.~\ref{Fig:NCSDriven} that a cooling depth of $L_{\rm cool}=242d\tau$ leads to a stable unit-step behavior in the topological charge. The behavior of the cooled $N_{\rm CS}(t)-N_{\rm CS}(0)$ for a selection of ten different initial conditions is given in the right panel of Fig.~\ref{Fig:NCSDriven}.

\begin{figure}
\centering
 \includegraphics[scale=0.33]{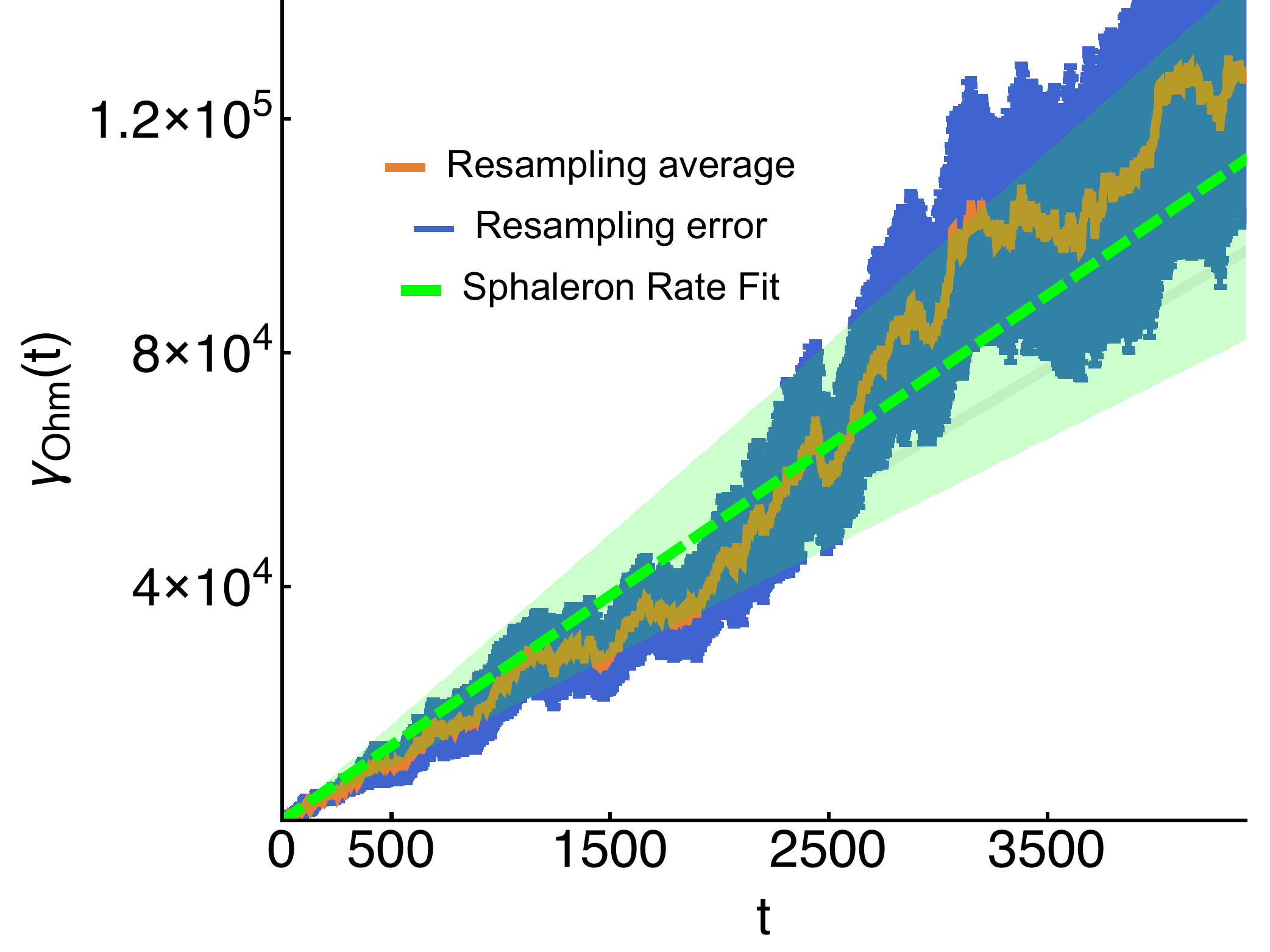}
 \includegraphics[scale=0.33]{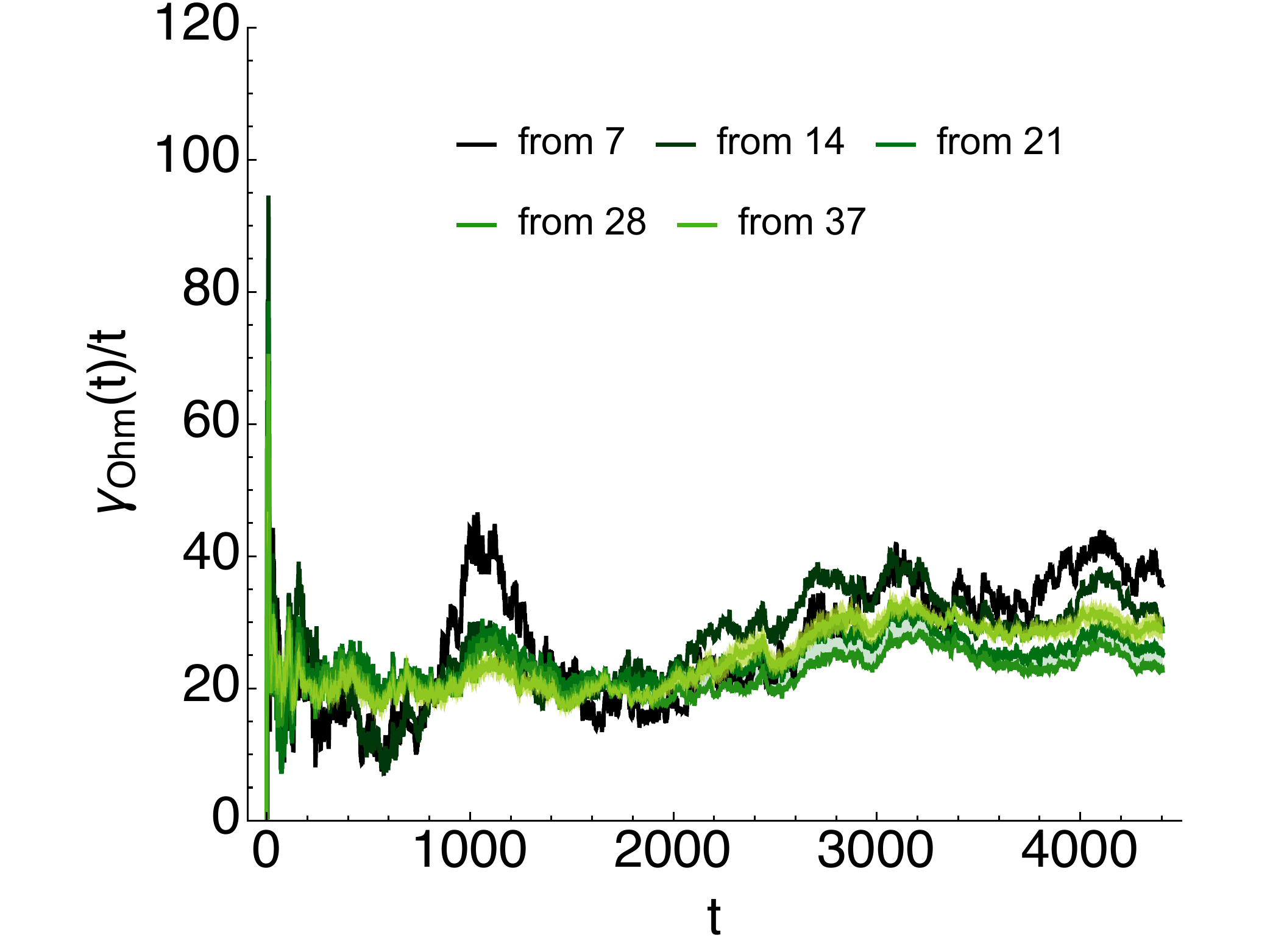}
 \caption{Numerical results of nonchiral Langevin theory: (Left) Estimating the sphaleron rate $\Gamma$ from a linear fit (green) of the resampled $\gamma(t)$ (orange). The resampling error estimate is given in blue and the gray curve denotes the value for $\Gamma=22.1$ observed in Ref.~\cite{Moore:2010jd}; (Right) The value of $\gamma(t)/t$ estimated from different number of ensemble realizations. }
 \label{Fig:SphaleronDriven}
\end{figure}

We estimate the value of the sphaleron rate in this scenario again from a linear fit to the resampled values of $\gamma(t)$ as shown in the left panel of Fig.~\ref{Fig:SphaleronDriven}. The systematic errors are assigned such that the fluctuating sample average (orange) lies within the $1 \sigma$ cone of the fit. The value we obtain here is 

\begin{align}
 \Gamma^{\rm Ohm}=26\pm7, 
\end{align}

\noindent which agrees with the result $\Gamma^{\rm Ohm}_{[75]}=22.10\pm 0.62$ in Ref.~\cite{Moore:2010jd} within our relatively large error bars. Note that the amount of cooling required in the driven case is actually less than for pure Hamilton dynamics and also the dependence of the sphaleron rate on the number of used ensemble members is weaker, as can be seen from the right panel of Fig.~\ref{Fig:SphaleronDriven}, where $\gamma(t)/t$ is shown.

\subsection{Langevin-type effective theory with chiral imbalance}

The genuinely new ingredient in this study is the inclusion of the chiral imbalance $n_5(t)$ as an effective but nevertheless explicit degree of freedom. Its back-reaction onto the gauge fields arises from the  anomalous current proportional to the magnetic field entering Eq.~\eqref{Eq:AnomalousEOM}. We start the real-time evolution of the system from the same initial configurations as before but at first $t<t_0=15$ with no chiral imbalance present. As we have seen in Sec.~\ref{SecYMEFT} this is necessary for the system to settle into the stationary state, where the electric energy of the dissipatively coupled system does not change with time anymore. At $t_0=15$ we then switch on $n_5(t_0=15)=n_0$ and observe the subsequent behavior of the system. Note that since the anomalous current is dissipationless the character of the stochastic noise has not changed and we thus find that the same cooling as in Sec.~\ref{SecYMEFT} with $L_{\rm cool}=272d\tau$ suffices for a robust determination of $N_{\rm CS}(t)-N_{\rm CS}(0)$.

\begin{figure}
\centering
 \includegraphics[scale=0.33]{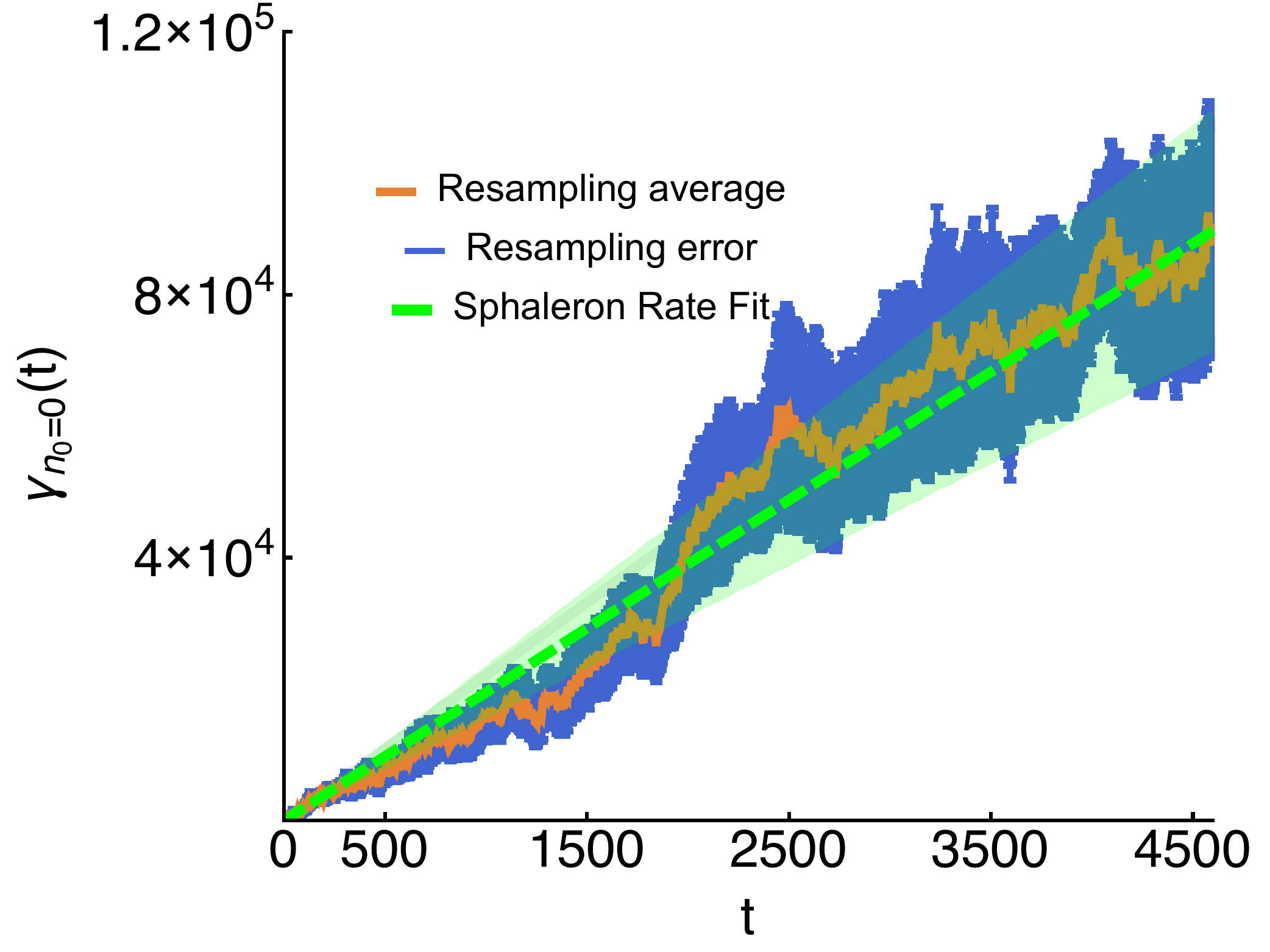}
 \includegraphics[scale=0.33]{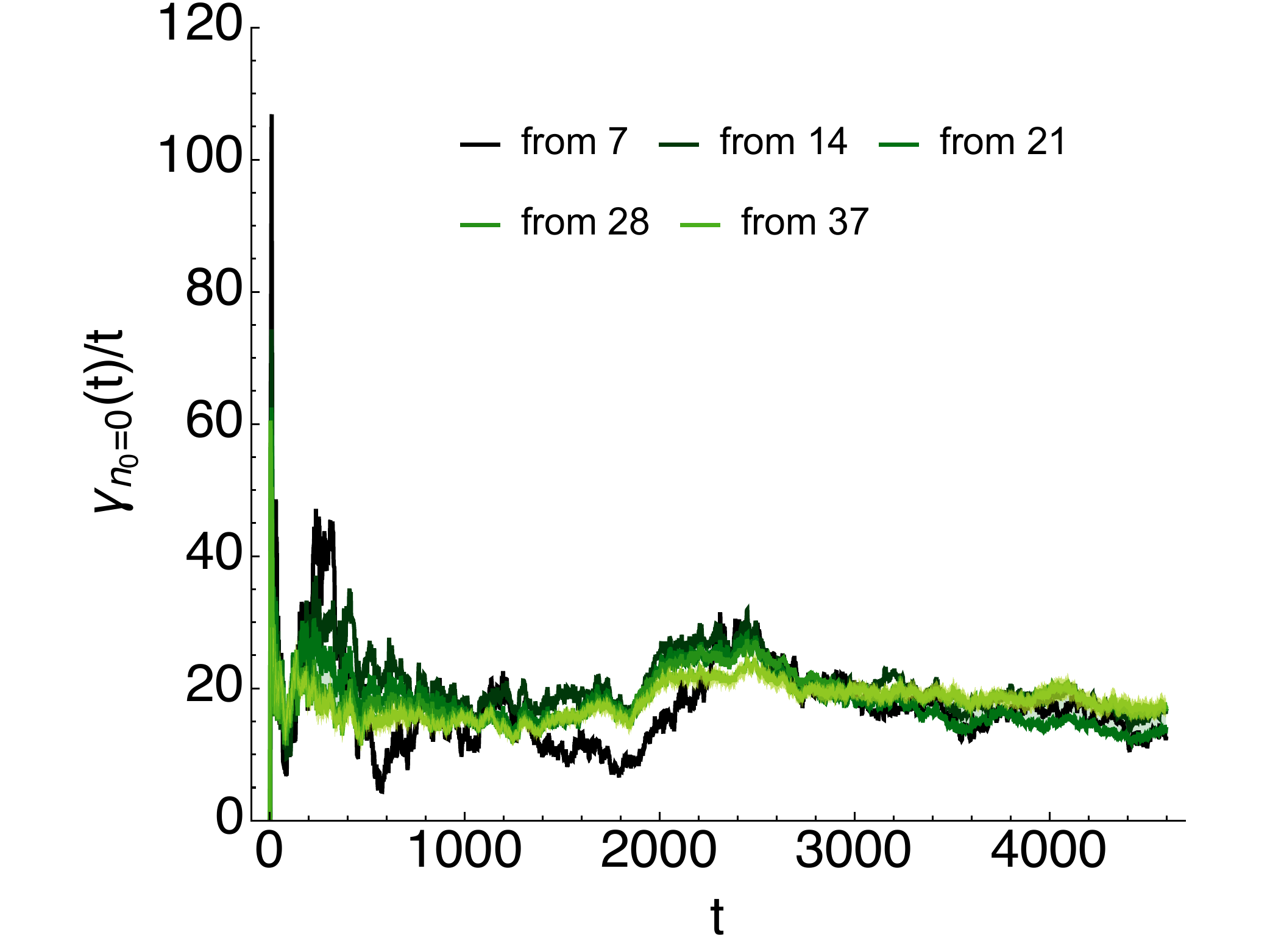}
 \caption{Numerical results of chiral Langevin theory with $n_0=0$: (Left) Estimating the sphaleron rate $\Gamma$ from a linear fit (green) of the resampled $\gamma(t)$ (orange). The resampling error estimate is given in blue and the gray curve denotes the value for $\Gamma=22.1$ observed in Ref.~\cite{Moore:2010jd} for the nonchiral Langevin theory; (Right) The value of $\gamma(t)/t$ estimated from different number of ensemble realizations.}
 \label{Fig:SphaleronN5OFF}
\end{figure}

The first question to ask is whether the interplay of the gauge sector with $n_5(t)$ via the anomalous current influences the sphaleron rate of the system even without a chiral imbalance present at first. To this end we compare the value of $\Gamma$ of Sec.~\ref{SecYMEFT}, $\Gamma^{\rm Ohm}=26\pm7$, with the one obtained with vanishing initial $n_5(t_0)=0$,
\begin{align}
 \Gamma^{\rm n_0=0}=20\pm4, 
\end{align}
as shown in Fig.~\ref{Fig:SphaleronN5OFF}.  Since the error bars are still relatively large, we cannot distinguish the two values statistically. It would however be interesting to reduce the uncertainty in a future study to understand whether this apparent agreement is accidental or a reflection of physics.

As discussed in the introduction section, analytic considerations have lead to the conclusion that in the presence of a chiral imbalance fluctuating gauge fields can develop instabilities which drive rapid dynamics in the magnetic sector at early times \cite{Akamatsu:2013pjd}. The back-reaction of the system on the unstable modes will eventually lead to a saturation of their energy and the system is expected to enter a new steady state.

\begin{figure}
\centering
 \includegraphics[scale=0.315]{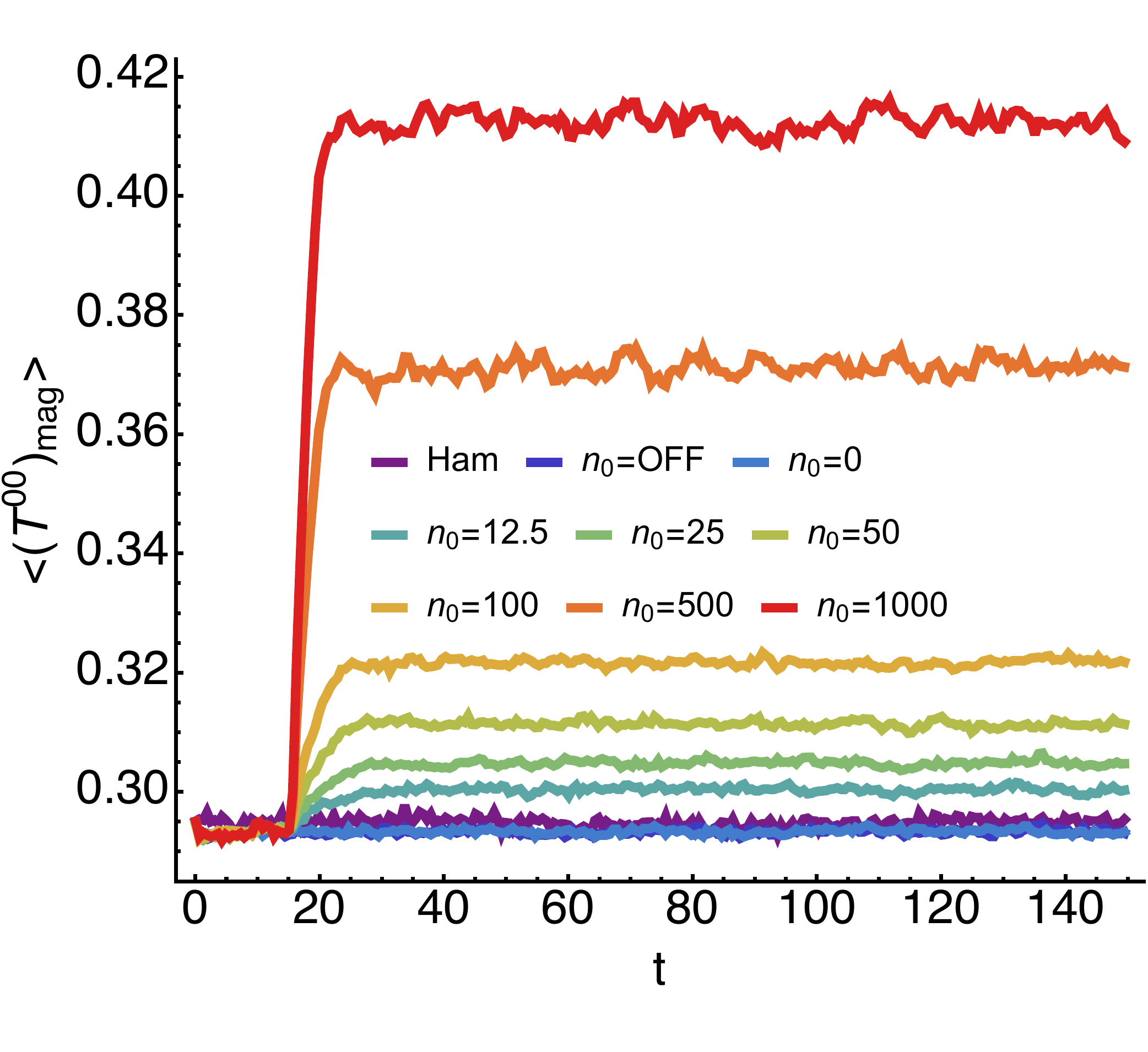}
 \includegraphics[scale=0.315]{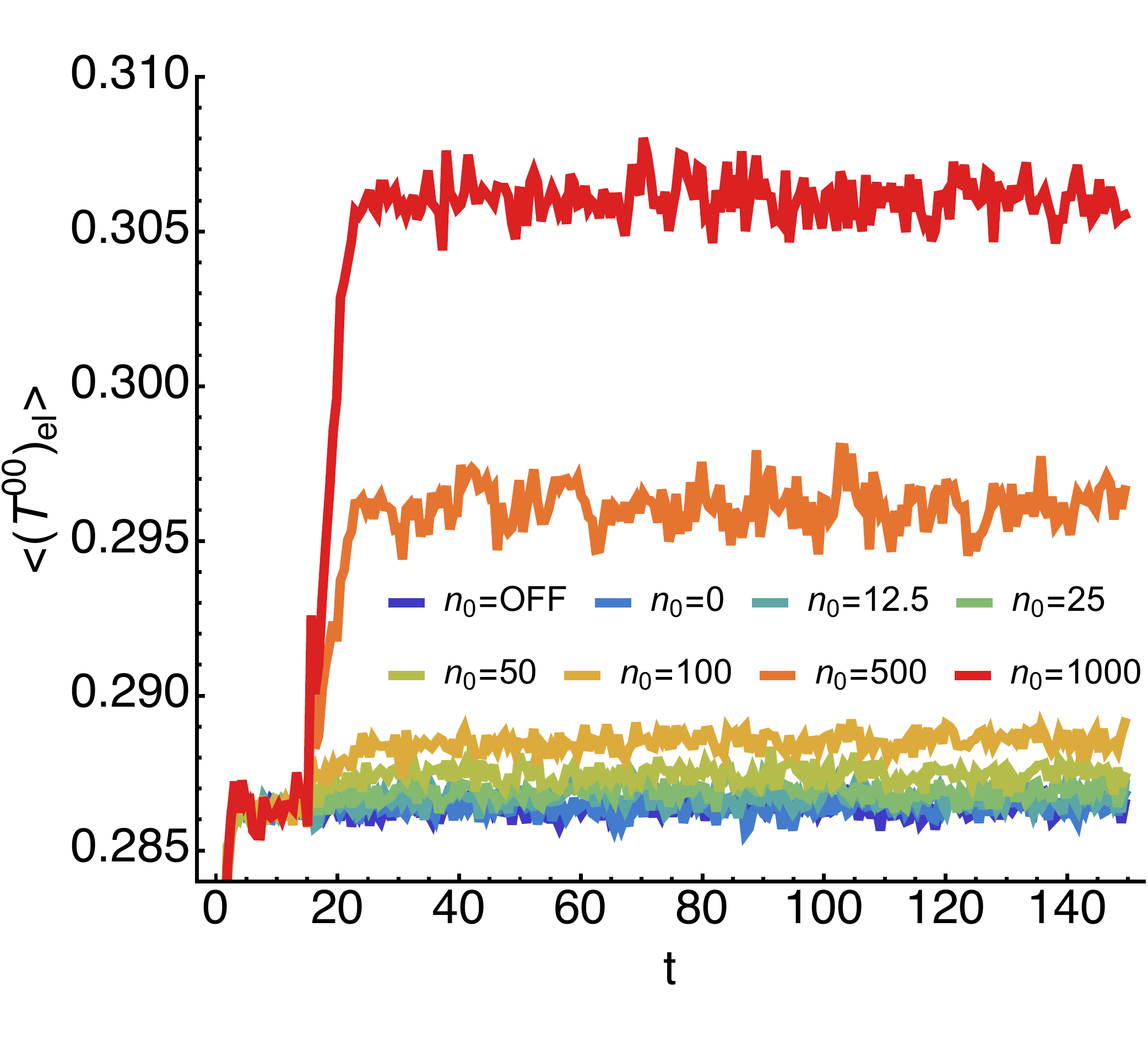}
 \caption{Numerical results of chiral Langevin theory: (Left) Magnetic energy density of the gauge fields at early times with the chiral imbalance switched on at $t_0=15$. One observes a rapid rise, clearly ordered with increasing values of $n_0$. A simple inspection by eye hints at three different regimes, a very brief but rapid one shortly after $t_0$ followed by a linear increase until saturation is reached; (Right) The corresponding electric energy density, which has stabilized at $t\simeq10$ from its initial rise before following the magnetic energy density with a rapid increase after $t_0$.}
 \label{Fig:EnergyAnomalous}
\end{figure}

In Fig.~\ref{Fig:EnergyAnomalous} we show the electric and magnetic energy of the system at early times for different values of $n_0=0,12.5,25,50,100,500,1000$. They indeed exhibit a rapid increase instantly after switching on the imbalance at $t_0=15$. Looking at the magnetic sector at intermediate values of $n_5$, it appears as if three different regimes can be identified. A very rapid early rise, followed by a slightly longer linear phase and eventually saturation to the new steady state. Not only does the value at which the energy saturates increase with larger values of $n_0$ is also reached a slightly earlier times indicating a more rapid instability at larger imbalances. The electric fields also show a rise over the same timescales, which however is smaller in magnitude. Interestingly the electric energy never overtakes its magnetic counterpart, an a posteriori justification of the assumption that $E_{\rm el}<E_{\rm mag}$ made in the derivation of the effective dynamics (see Tab.~\ref{tab:softscales1}).

\begin{figure}
\centering
 \hspace{-0.45cm}\begin{minipage}{6in}
  \centering
  \raisebox{-0.5\height}{\includegraphics[scale=0.3]{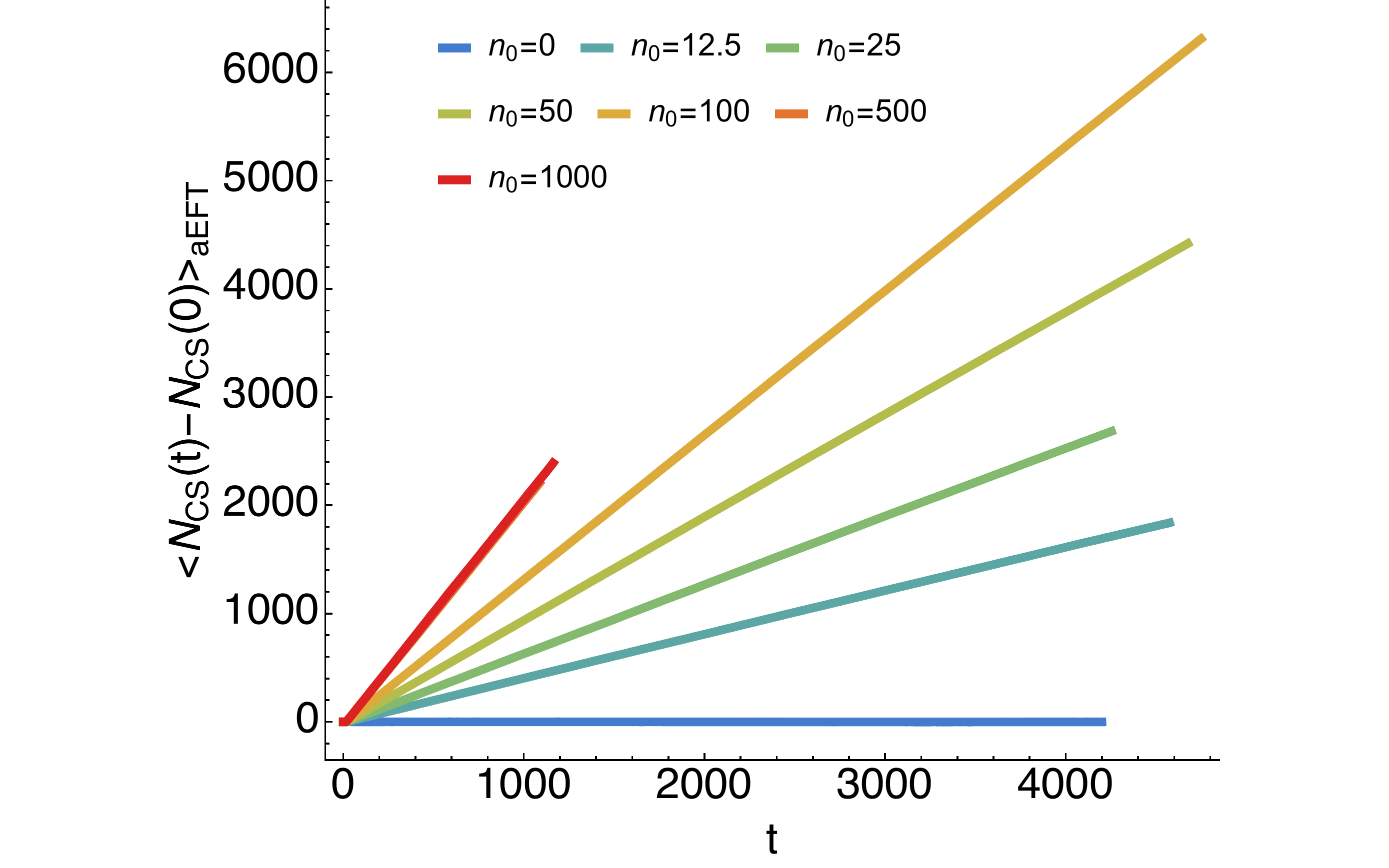}}
  \hspace*{.0in}
  \raisebox{-0.5\height}{ \includegraphics[scale=0.3]{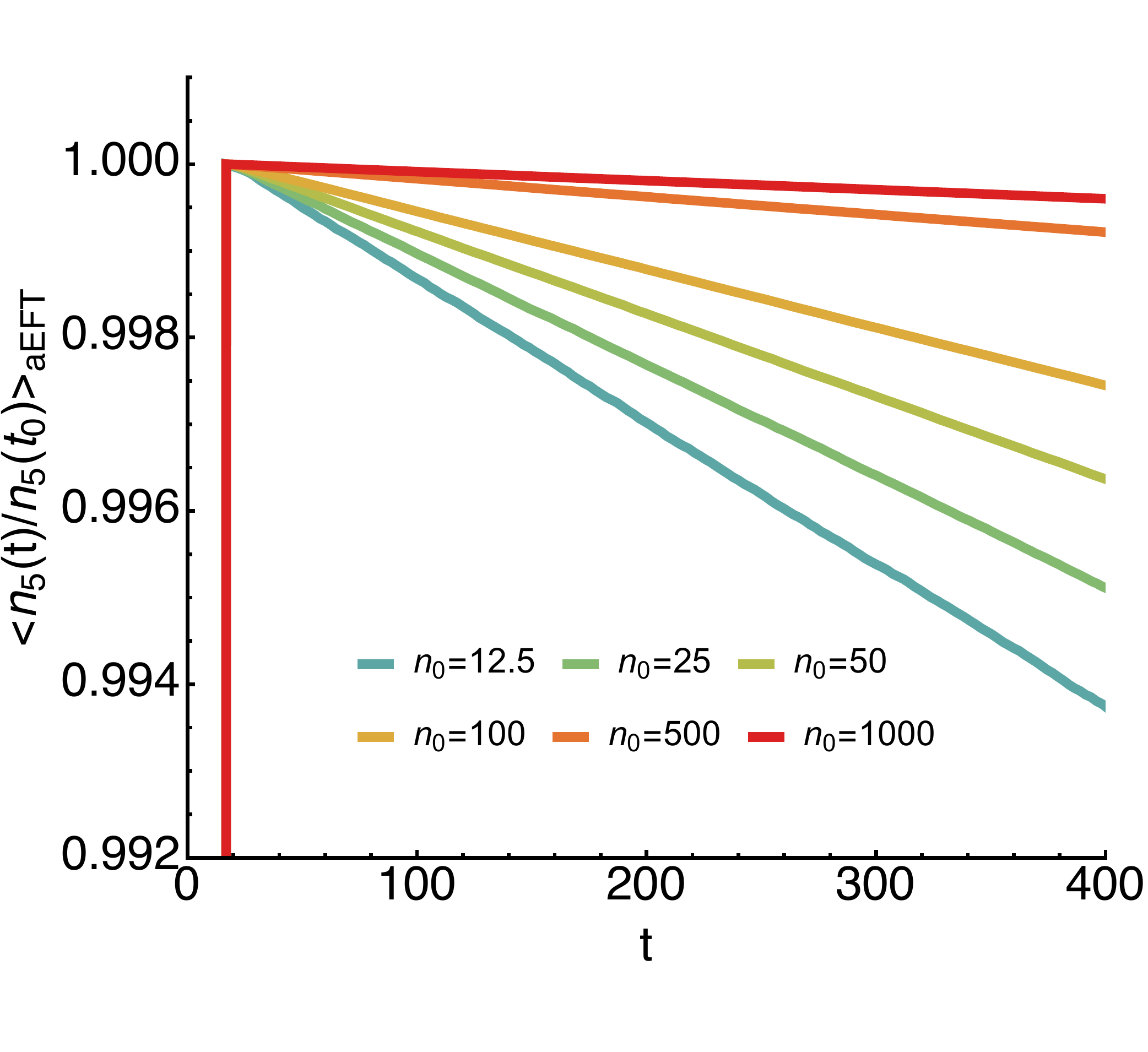}}
\end{minipage} 
 \caption{Numerical results of chiral Langevin theory: (Left) The averaged $\langle N_{\rm CS}(t)-N_{\rm CS}(0) \rangle$ measured for different values of $n_0=0,\ldots,1000$. A clear ordering in the slope of the linear rise according to the value of $n_0$ is observed. For a too rapid rise $n_0\gtrsim100$ temporal resolution does not suffice to accommodate all transitions, which leads to an unphysical upper limit of the slope; (Right) Evolution of the chiral imbalance $n_5(t)$ normalized to unity at $t_0=15$. Due to helicity conservation the increase in values of $\langle N_{\rm CS}(t)-N_{\rm CS}(0) \rangle$ signal its decrease towards zero over the same time span.}
 \label{Fig:NCSAnomalous}
\end{figure}

Let us have a look at the behavior of the topological charge in the presence of a chiral imbalance. As an example we showcase $N_{\rm CS}(t)-N_{\rm CS}(0)$ at $n_0=25$ in the left panel of Fig.~\ref{Fig:NCSAnomalous}. Before $t_0$ the same changes in topology occur as in the case without anomalous current but once $n_5$ is present a clear drift among $N_{\rm CS}(t)-N_{\rm CS}(0)$ is induced. The drift is topology driven, as it occurs via well separated transitions of unit steps, which when taken together lead to a linear increase of the topological charge over time. For larger values of $n_5$ even more transitions per time occur, so that eventually our temporal resolution will not be able to accommodate all of them and we will end up with a saturation of the drift. This is exactly what we find in the left panel of Fig.~\ref{Fig:NCSAnomalous}, where the averaged $\langle N_{\rm CS}(t)-N_{\rm CS}(0) \rangle$ is shown. Our choice of $a_t=0.0375$ appears to give reliable results up to $n_0=100$, beyond which saturation effects become significant.

\begin{figure}
\centering
 \includegraphics[scale=0.31]{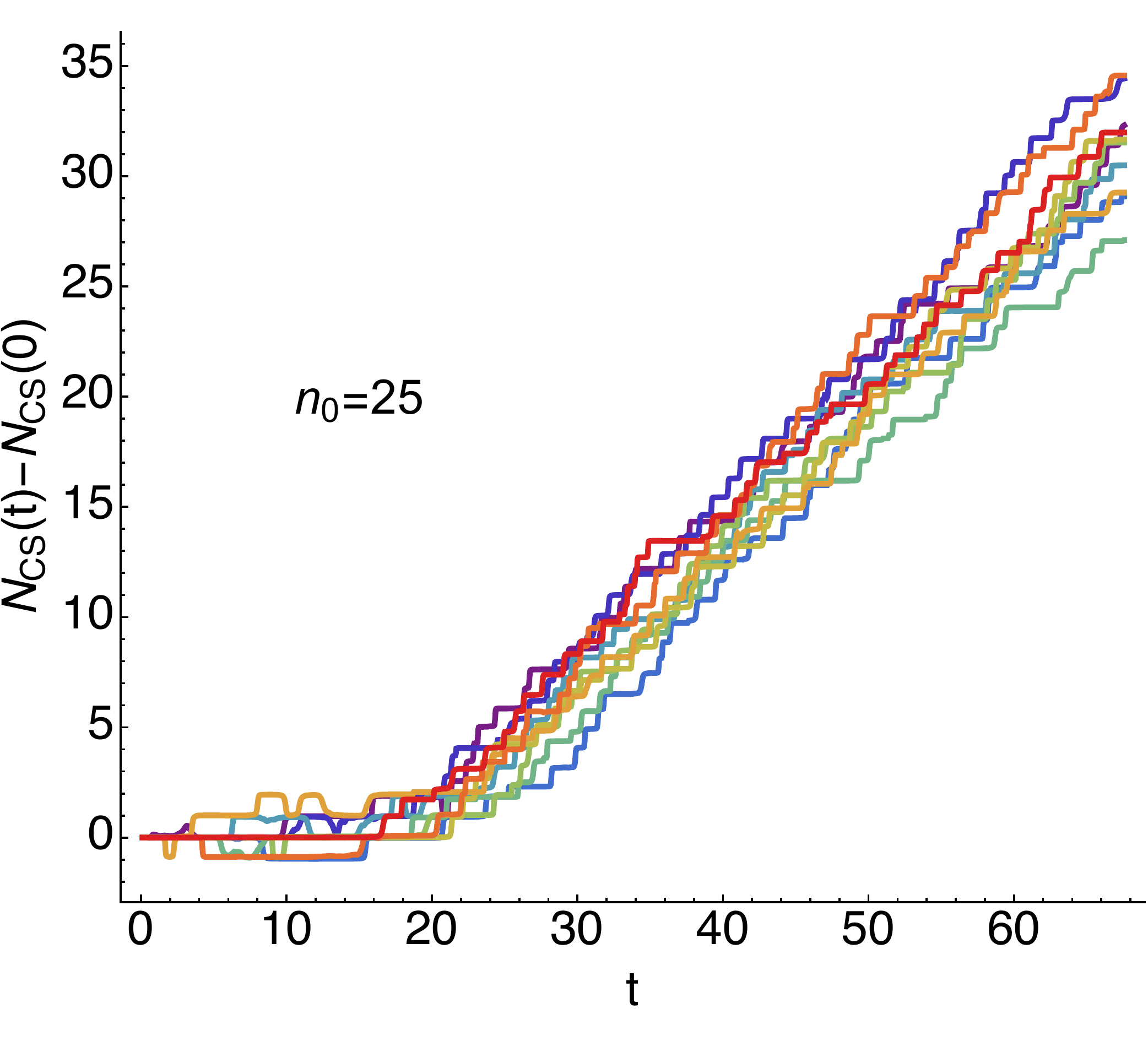}
 \includegraphics[scale=0.33]{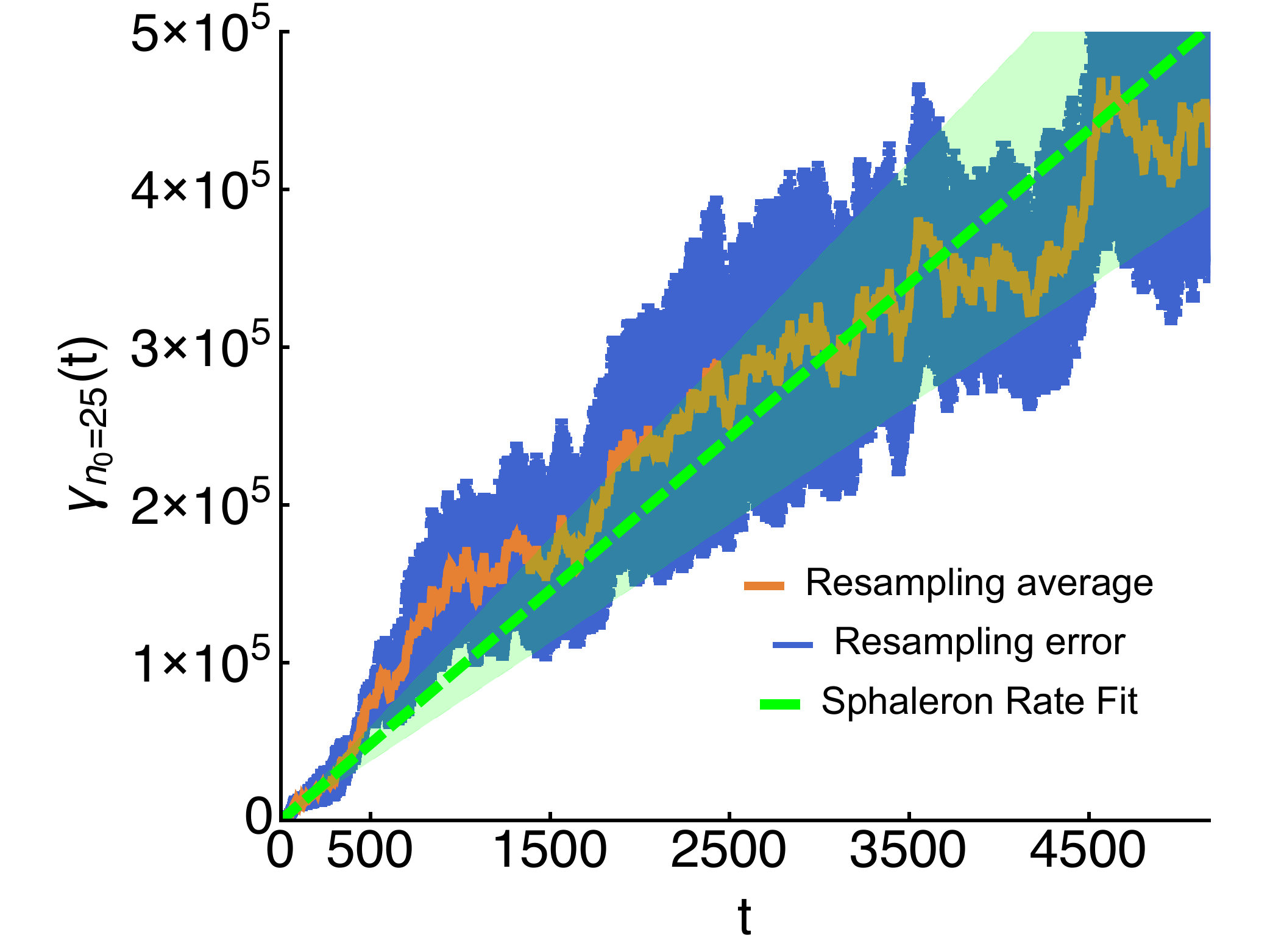}
 \caption{Numerical results of chiral Langevin theory with $n_0=25$: (Left) Early time behavior of $N_{\rm CS}(t)-N_{\rm CS}(0)$ after cooling, starting from a selection of ten different thermal initial configurations. With the presence of a finite $n_5(t)$ a clear drift to larger values is observed, which proceeds via well separated transitions between individual vacuum sectors. The increase if averaged over different realizations appears linear. (Compare Fig.~\ref{Fig:NCSAnomalous}); (Right) Estimating the sphaleron rate $\Gamma$ from the slope of $\gamma(t)$. We resample twenty times by drawing a subset of nineteen ensemble realizations from which the average (orange) and variance (blue) of $\gamma(t)$ follows. Assuming a diffusive process we fit with a linear function without intercept (green).}
 \label{Fig:NCSAnomalousIndiv}
\end{figure}

One expects that the induced changes in the gauge field will diminish the chiral imbalance, in line with the requirement of helicity conservation. We have explicitly checked it to be fulfilled in our numerical implementation. And indeed the increase in $N_{\rm CS}(t)-N_{\rm CS}(0)$ thus leads to a decrease of $n_5(t)$ over time as can be seen on the right of Fig.~\ref{Fig:NCSAnomalous}. It is our particular choice of $\sigma_c=1$, $\beta_L$ and $n_0$ that yields a rather slow decrease in $n_5(t)$ here. Interestingly, when rescaled to unity at $t_0$ the relative decrease is weaker, the larger the initial $n_0$ is. 

While currently too costly, an investigation of the very late time behavior of the system would be highly desirable. In particular it would be interesting to observe whether, the chiral imbalance is fully depleted in the end and to what state the system relaxes if $n_5=0$ is eventually reached. As we saw that the system is stable for $n_0=0$ and no replenishment of the imbalance occurs spontaneously, once $n_5(t)$ is depleted, we expect to eventually return to a state of vanishing drift, diffusing around a large but constant value of the topological charge.

\begin{figure}
 \centering
 \hspace{-0.45cm}\begin{minipage}{6in}
  \centering
  \raisebox{-0.5\height}{\includegraphics[scale=0.34]{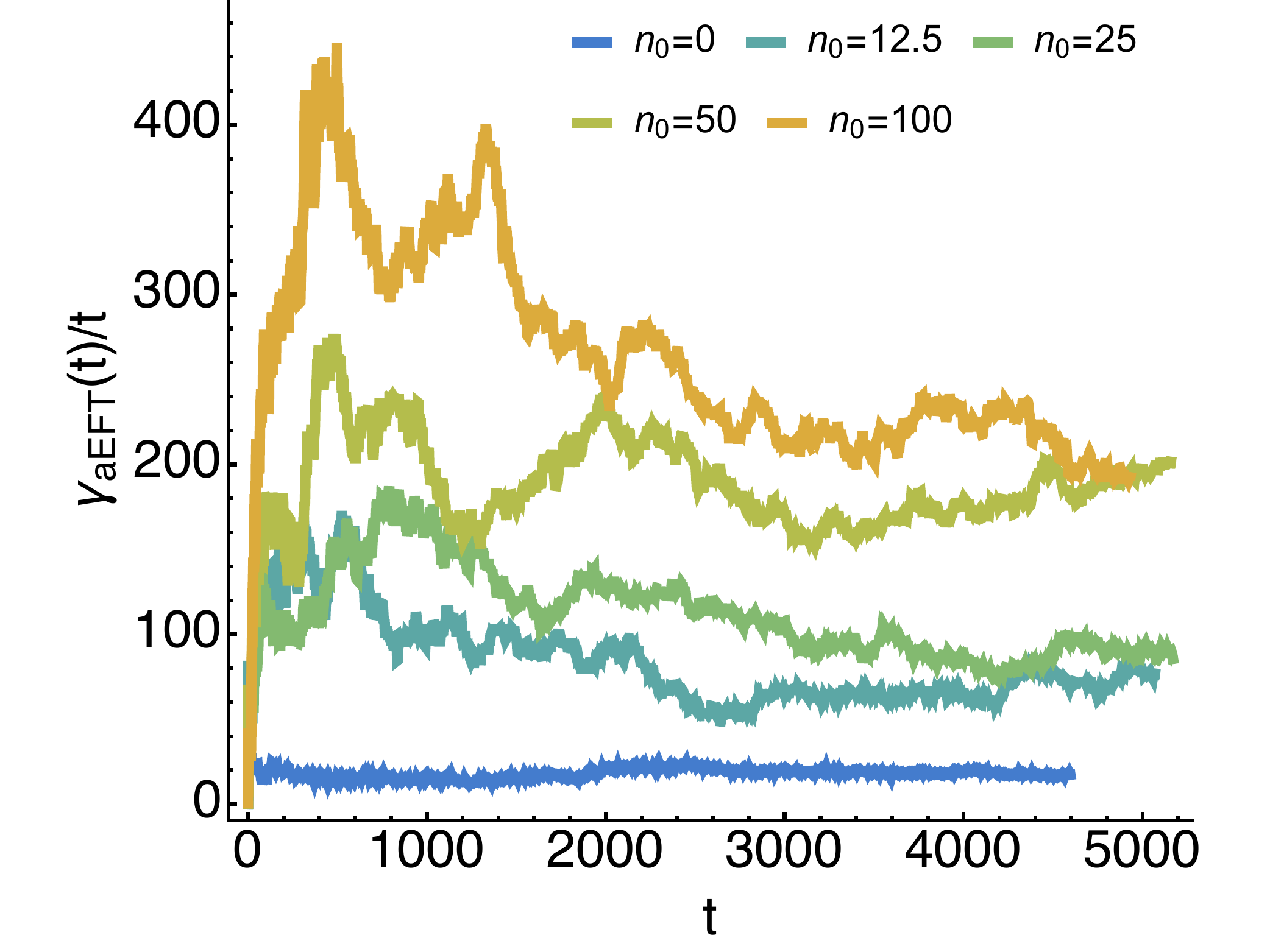}}
  \hspace*{.2in}
  \raisebox{-0.5\height}{ \includegraphics[scale=0.32]{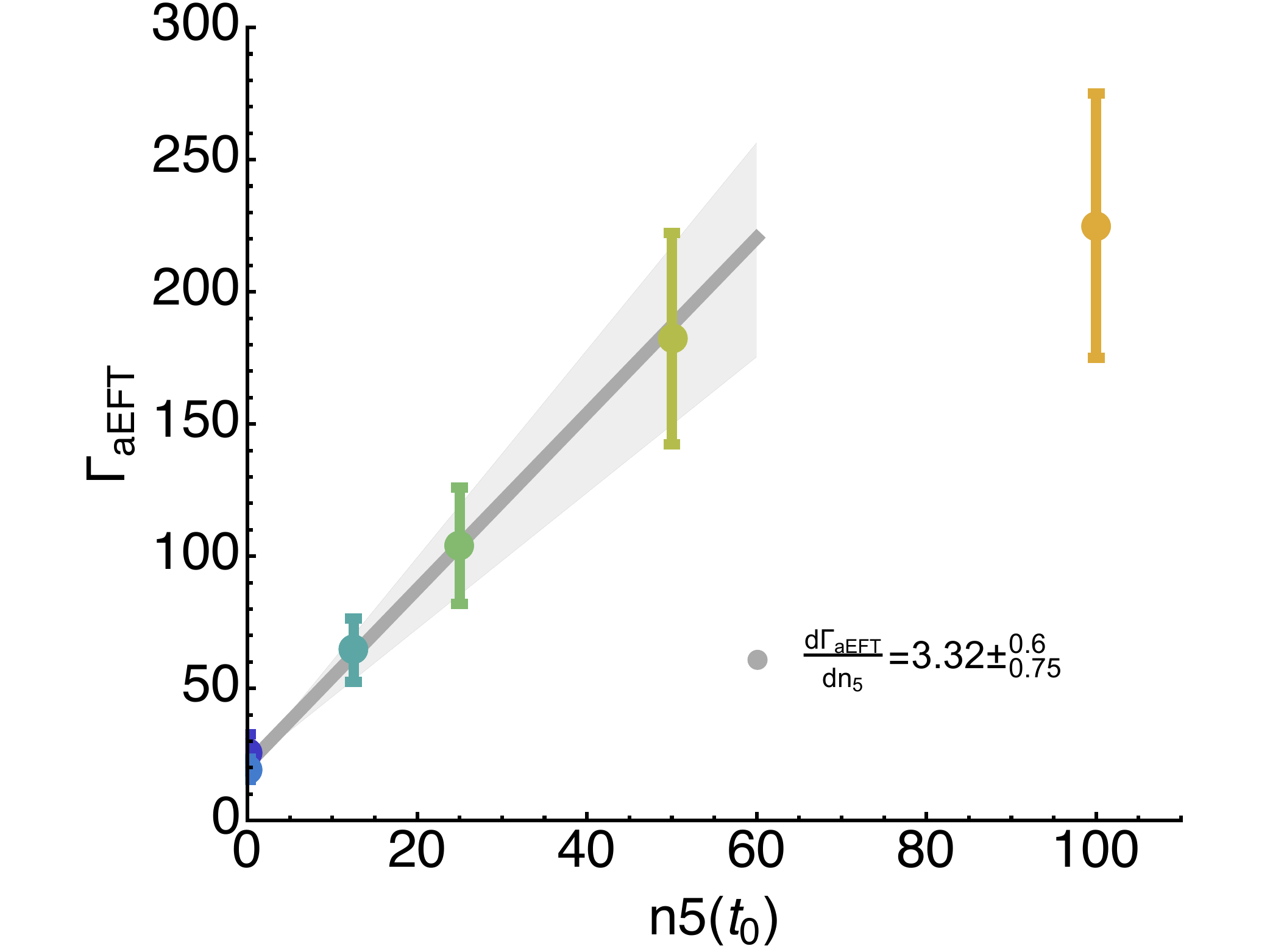}}
\end{minipage} 
 \caption{Numerical results of chiral Langevin theory: (Left) The sphaleron rate for different values of initial chiral imbalance. The late time values $\Gamma$ show a clear ordering with respect to $n_0$, which if plotted (right) appears to show a linear relation. Note that we subtract the drift of the average $\langle N_{\rm CS}(t)-N_{\rm CS}(0) \rangle$ from the sphaleron rate.  }
 \label{Fig:SphaleronAnomalous}
\end{figure}

Now we can proceed to inspect the sphaleron rate in the presence of a chiral imbalance, the relevant plots are given in Fig.~\ref{Fig:SphaleronAnomalous}. For completeness we plot the values for $\gamma(t)/t$ in the left panel to qualitatively illustrate that there is an ordering in the sphaleron rate with $n_0$. The actual values of $\Gamma$ are obtained as before from linear fits to $\gamma(t)$, summarized in Tab.~\ref{Tab:SphalaEFT} and plotted in the right hand pane of Fig.~\ref{Fig:SphaleronAnomalous}. 

\begin{table}[h!]
\centering
 \begin{tabular}{|c|c|c|c|c|c|c|}
  \hline
  ~& Ohmic &$n_0=0$ & $n_0=12.5$ & $n_0=25$ & $n_0=50$ & $n_0=100$ \\
  \hline
  $\Gamma$ & $26\pm7$ & $20\pm4$ & $65\pm12$ & $104\pm22$ & $182\pm40$ & $225\pm50$\\ \hline
  \end{tabular}\caption{The estimated values for the sphaleron rate in the presence of a chiral imbalance $n_0$.}\label{Tab:SphalaEFT}
\end{table}

While it is not easy to distinguish from the left panel of Fig.~\ref{Fig:SphaleronAnomalous}, the values for $\Gamma$ obtained by the fits are well described by a linear dependence with slope
\begin{align}
\frac{d \Gamma}{dn_0} = 3.32 \pm ^{0.6}_{0.75} \end{align}
up to $n_0=50$. Note that for $n_0=100$ the extracted sphaleron rate lies significantly below this trend. A reason for this deviation might be that at $n_0=100$ the drift in $N_{\rm CS}(t)-N_{\rm CS}(0)$ has become too fast to be adequately resolved by our temporal step size. I.e. it is related to the saturation of $N_{\rm CS}(t)-N_{\rm CS}(0)$ itself, observed in Fig.~\ref{Fig:NCSAnomalous}.

As all previous results have been obtained at a single lattice spacing we need to make sure that the phenomena we observe are not simply artifacts of the finite lattice spacing discretization. To probe the small spatial lattice spacing limit within a classical statistical simulation in thermal equilibrium we can increase $\beta_L$. In order to make sure that we retain the relevant physics of the infra-red regime, the lattice volume has to increase at the same time so that $\beta_L/N_s={\rm const.}$ 

\begin{figure}
 \centering
\hspace{-0.45cm}\begin{minipage}{6in}
  \centering
  \raisebox{-0.5\height}{\includegraphics[scale=0.36]{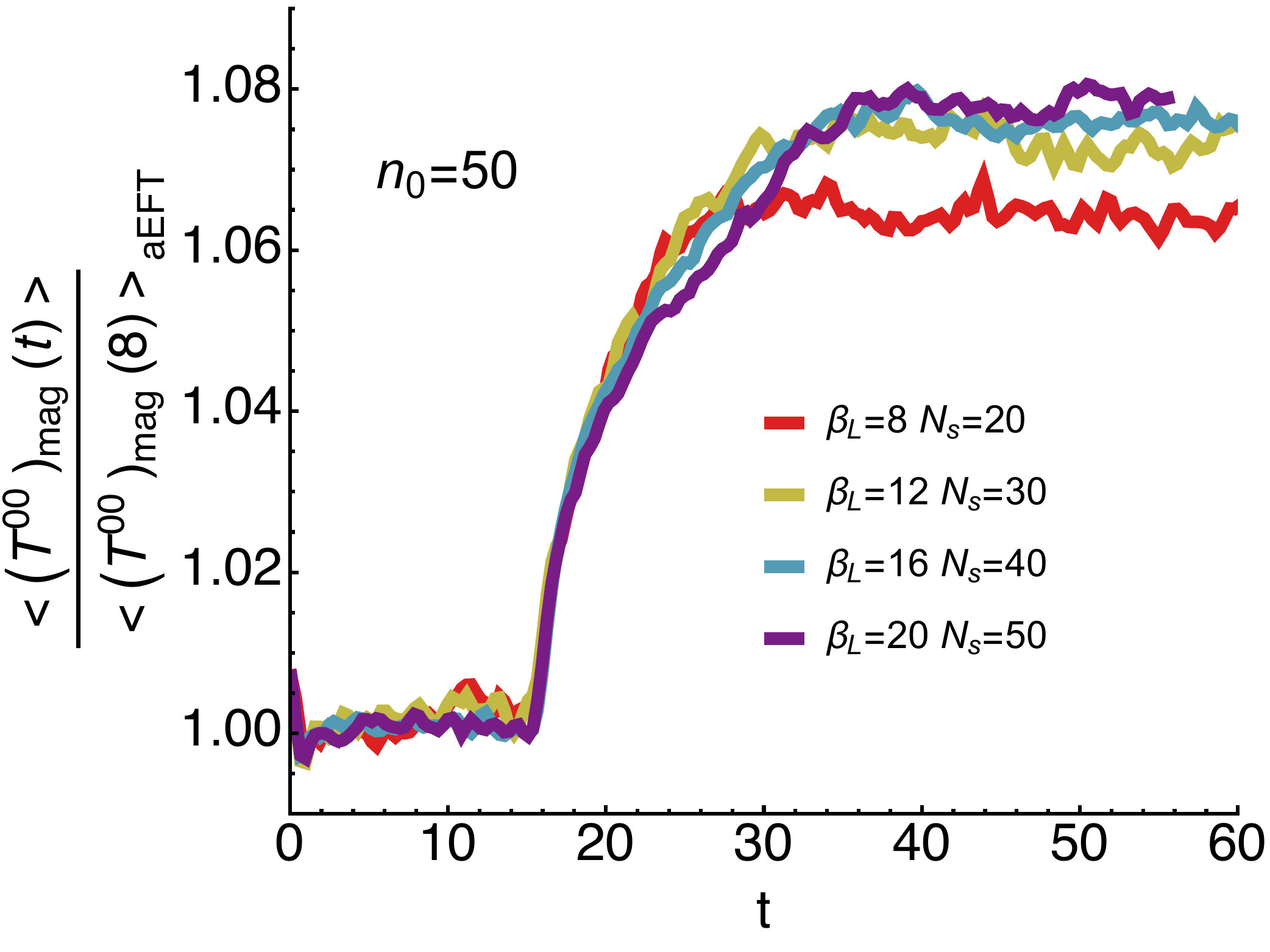}}
  \hspace*{.2in}
  \raisebox{-0.5\height}{ \includegraphics[scale=0.315]{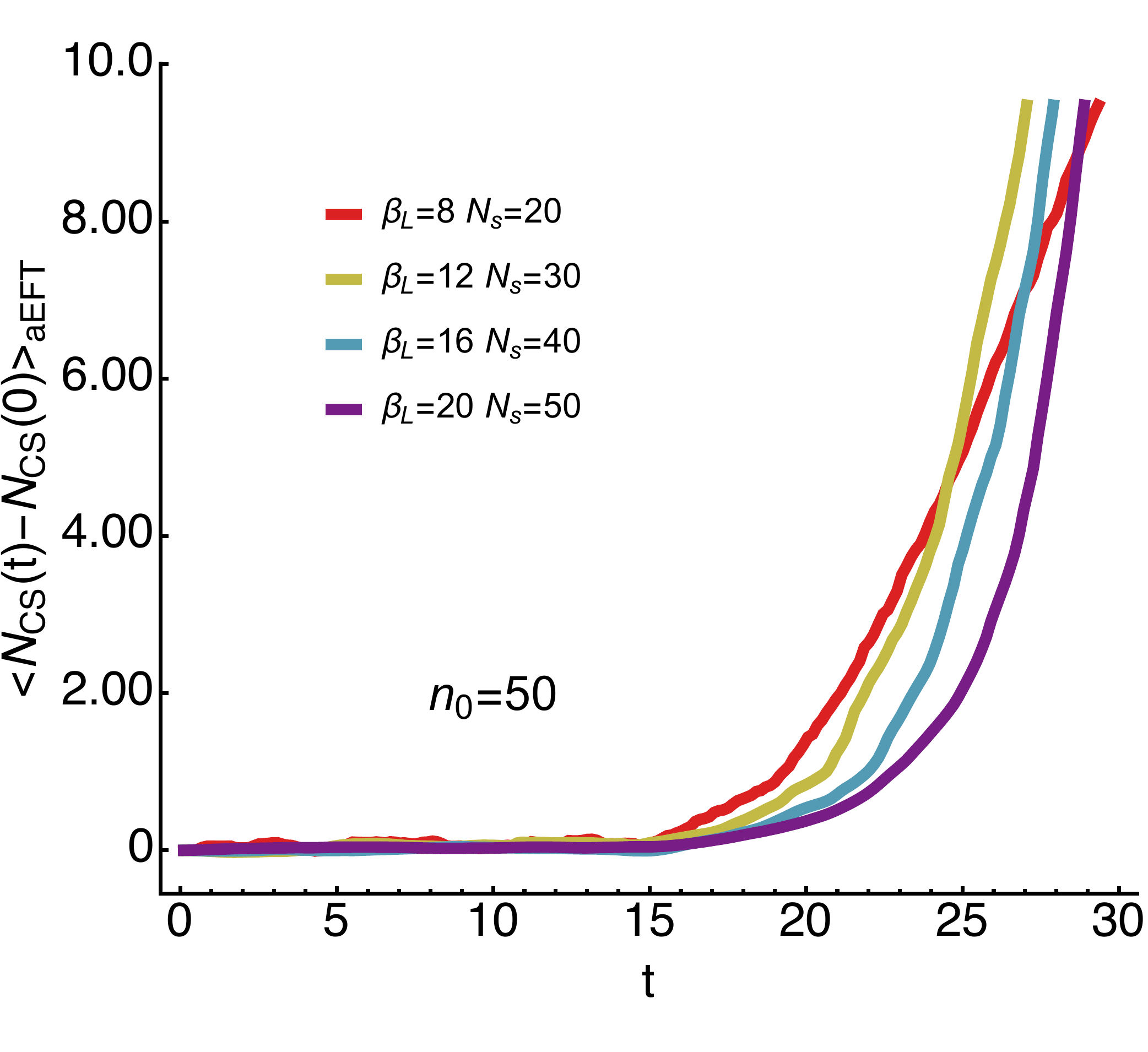}}
\end{minipage} 
\caption{Numerical results of chiral Langevin theory with $n_0=50$: (Left) Time evolution of the normalized magnetic energy for different $\beta_L$ in a fixed physical volume $\beta_L/N_s=0.4$. The relative energy at which the instability saturates increases slightly with $\beta_L$ but already at $\beta_L=20$ seems to have reached its continuum value. Note that the absolute value of the energy decreases with $\beta_L$. The onset of the instability at $t_0=15$ is not affected by the smaller lattice spacing; (Right) Average of the topological charge for different $\beta_L$.  The set-in of the drift shifts to slightly larger times  but the early time region during the instability between $t=15-20$ shows only minimal change between $\beta_L=16$ and $\beta_L=20$. Note that the slope of the linear rise becomes steeper with larger $\beta_L$ but seems to saturates at around $\beta_L=20$.
}
 \label{Fig:CompDiffBetaAnomalous}
\end{figure}

In the normalized magnetic energy shown on the left of Fig.~\ref{Fig:CompDiffBetaAnomalous} we find that as the chiral imbalance is switched on, the onset of initial fast dynamics occurs independent of $\beta_L$. The relative value at which the instabilities saturate on the other hand grows slightly with $\beta_L$ but seems to saturate already at $\beta_L=20$. In absolute terms this limiting value decreases, as the energy in the system is diminished by lowering the temperature.  

Looking at the average of the Chern-Simons number on the right panel tells us that the onset of the linear rise seems to shift slightly to later values. The slope becomes steeper as we go from $\beta_L=8$ to $\beta_L=20$ but it appears to have saturated at the larger value. Note that we have not adapted the cooling prescription for different $\beta_L$, so that a quantitative interpretation of the drift in Fig.~\ref{Fig:CompDiffBetaAnomalous} should not be attempted. What we can however infer is that the phenomenon of chiral instabilities survives in the $a_s\to0$ limit.

The other limit concerns the temporal step size. Unfortunately in the context of this study we do not have the computational resources to perform a thorough analysis of the $a_t$ dependence of the late time sphaleron rate.  On the other hand, according to the findings of Ref.~\cite{Moore:2010jd} the difference in $\Gamma$ between $a_t=0.1$ and $a_t=0$ is below $2\%$. This gives us confidence that with $a_t=0.0375$ the time discretization error is insignificant compared to the other systematic uncertainties.

\section{Discussion and conclusion}
\label{SecConcl}

We started out with two sets of questions. The first one concerned the existence and early-time evolution of the \textit{chiral instability}. The second one aimed at understanding the \textit{fate of fermion chiral imbalance}: how $n_5$ and $N_{\rm CS}$ behave at late times. For both cases this exploratory numerical study has provided qualitative insight.

The basis for our simulation of the real-time evolution is a chiral Langevin theory \cite{Akamatsu:2014yza} in which the soft gauge sector is coupled to hard chiral fermions via both an Ohmic current, as well as a chiral magnetic current. The interaction between the chiral imbalance and the topology of the gauge sector is dynamically included. 
We have tested the numerics of cooling, required for the lattice determination of $N_{\rm CS}$, as well as those for solving Hamilton equations of motion in the pure Yang-Mills sector for $N_s=20$ and $\beta_L=8$.  While the values we obtain for the diffusion of topological charge lie quite close to those reported in Ref.~\cite{Moore:1998swa}, we note that our results are systematically smaller. The reason for this deviation might be related to our choice of fully cooling to the vacuum at each time-step, the use of the standard clover approximation for the field strength tensor and a different number of ensemble realizations used.  On the other hand our sphaleron rate in the driven system without anomalous effects, agrees within $1\sigma$ with the corresponding reference value obtained in Ref.~\cite{Moore:2010jd}.

In the chiral Langevin theory we find that for an initially depleted $n_5$ the system is stable and on average no chiral imbalance is generated. The system diffuses around the topological sector it started from and shows within errors the same diffusion constant as if the anomalous term is switched off by hand. Increasing the statistics for these two cases in the future is highly desirable, as it will allow us to reduce the still relatively large error bars on the sphaleron rate. In turn we will be able to settle, whether the $1 \sigma$ difference between the Ohmic and $n_0=0$ rate diminishes or indeed different diffusion occurs. We actually expect that the picture of conventional diffusion of Chern-Simons number needs to be modified even for $n_0=0$. Once an individual realization of the system evolves for long enough times that changes in topological charge and thus fermion chiral imbalance can accumulate, the effects of the anomaly can become relevant and reduce the growth of $\langle (N_{\rm CS}(t)-N_{\rm CS}(0))^2\rangle$. The systematically smaller central value of the sphaleron rate in the presence of the anomalous term compared to the purely Ohmic theory might be an indication of this mechanism.

In the presence of a chiral imbalance, we find that modes in the soft gauge fields indeed become unstable. A \textit{chiral instability} ensues with a rapid increase of magnetic and electric energy, which is fed by the hard sector. For intermediate values of initial $n_5(t_0)$, an inspection by eye hints at the existence of three different regimes. A very fast initial rise of $T^{00}_{\rm mag}$ is followed by a slower linear increase that saturates to an apparent constant. Note that $n_5$ at that time is still appreciably different from zero and thus will continue to feed topology changing processes in the gauge sector. Conservation of energy in the overall system implies that further energy should be accumulated in the gauge fields. Since for our choice of parameters however the decrease of $n_5$ proceeds very slowly (see Fig.~\ref{Fig:NCSAnomalous}) the accompanying minute changes in energy might still be masked by statistical noise. The value of energy at which the instability saturates increases with the size of the initial imbalance $n_5(t_0)$. Not only is more energy deposited in the gauge fields but the time it takes to reach the saturation also diminishes. When decreasing the spatial lattice spacing towards zero by increasing $\beta_L$ we find that the instability prevails and hence is not just an artifact of discretization.

Interestingly the presence of a finite $n_5$ results in a clear linear drift over time in the average Chern-Simons number. Inspecting the individual ensemble realization affirms its topological nature, as it is composed of well separated unit step transitions (see. Fig.~\ref{Fig:NCSAnomalousIndiv}). A deviation of $\langle N_{\rm CS}(t)-N_{\rm CS}(0)\rangle$ from zero is not surprising if $n_5\neq0$. The linear time dependence exhibited here on the other hand is puzzling, as the non-Abelian character of the gauge fields does not immediately suggest such a simple relation. Actually if we consider as model of the evolution of $N_{\rm CS}$ and $n_5$ a simple coupled random walk, the linear dependence on the initial value is just a manifestation of the early time rise of an exponential behavior that will eventually asymptote to its equilibrium value. Comparing the rise in energy (Fig.~\ref{Fig:EnergyAnomalous}) and the changes in topology (Fig.~\ref{Fig:NCSAnomalousIndiv}) we see that the onset of the drift is located just after the saturation of the rise in the magnetic and electric field energy. 

Due to helicity conservation the changes in $N_{\rm CS}$ act opposite to the imbalance by which they are fed. This reduces $n_5(t)$ monotonously. Even though we did not have the computational resources to follow the time evolution up to the full depletion of $n_5$, the stability of the system in the absence of $n_5$ tells us that eventually the drift in $N_{\rm CS}$ will abate and we do not expect a subsequent spontaneous replenishment of $n_5$ to occur: The linear rise observed at early time cannot be maintained forever and the system will begin to diffuse around a different topological sector than the one it started from. Whether the sphaleron rate at these times reverts to the one expected for the $n_5(t_0)=0$ case needs to be ascertained in a future study.

At intermediate times $(t=100-4500)$, i.e., after the saturation of the initial instability but before the value of $n_5$ has changed appreciably we find that the sphaleron rate shows a clear dependence on $n_5(t_0)$.  Once the initial imbalance becomes too large and the drift between topological sectors becomes too rapid to be resolved by our temporal spacing, a saturation of the rate is expected and also observed ($n_5(t_0)\geq100$). For smaller values of the initial imbalance we find a clear linear dependence of $\Gamma$ on $n_5(t_0)$. The interpretation of this finding and its possible phenomenological consequences in the context of heavy-ion collisions and early universe baryogenesis are work in progress.

Rapid developments in the simulations of nonequilibrium systems in the classical statistical regime bodes well for further insight into the questions investigated in our exploratory study. On the one hand state-of-the-art simulations (see, e.g., Ref.~\cite{Berges:2013fga}) now incorporate by default the expanding geometry of the heavy-ion collision region and make close contact with the saturation picture for the incident nuclei. The fact that the gauge field modes do not fill up the spectrum up to the cutoff in these simulations contrary to the thermal case, makes them systematically better controlled. On the other hand developments are underway to further develop the treatment of explicit fermionic degrees of freedom from first principles \cite{Kasper:2014uaa,Hebenstreit:2013baa,Gelis:2015eua,Hebenstreit:2013qxa,Borsanyi:2008eu} and to eventually include quantum corrections to classical statistical simulations. Progress in those directions most surely will provide us with new tools for understanding anomalous effects at early times.

\acknowledgments 
A.~R. thanks M.~Mace, S.~Schlichting and R.~Venugopalan for fruitful discussions. Insightful discussions on optimal dissipative cooling with D.~Sexty and I.-O.~Stamatescu and on early time particle production with N.~Tanji are gladly acknowledged. Numerical simulations were performed on the in-house cluster of the ITP at Heidelberg University and the used source code is available for public inspection at www.scicode.org/aEFT. 
Y.~A. is supported, in part, by JSPS Postdoctoral Fellowships for Research Abroad.
The work of N.~Y. was supported, in part, by JSPS KAKENHI Grants No. 26887032 and MEXT-Supported Program for the Strategic Research Foundation at Private Universities, ``Topological Science'' (Grant No. S1511006).

\end{document}